\documentclass[10pt]{article}
\usepackage{bbm}
\usepackage[final]{graphicx}
\usepackage{amsmath,amscd}
\usepackage{amsfonts,amsbsy}
\usepackage{amssymb}
\usepackage{tikz-cd}
\usepackage{upgreek}
\usepackage{subcaption}
\usepackage[utf8]{inputenc}
\usepackage{jheppub}
\usepackage{hyperref}
\setcitestyle{square}
\usepackage{bm}

\def\empile#1\over#2{\mathrel{\mathop{\kern 0pt#1}\limits_{#2}}}
\def\bs{\boldsymbol}

\newcommand{\slv}{\raise.15ex\hbox{$/$}\kern-.53em\hbox{$v$}}
\newcommand{\slF}{\raise.15ex\hbox{$/$}\kern-.53em\hbox{$F$}}
\newcommand{\slL}{\raise.15ex\hbox{$/$}\kern-.53em\hbox{$L$}}
\newcommand{\slP}{\raise.15ex\hbox{$/$}\kern-.53em\hbox{$P$}}
\newcommand{\slp}{\raise.15ex\hbox{$/$}\kern-.53em\hbox{$p$}}
\newcommand{\slq}{\raise.15ex\hbox{$/$}\kern-.53em\hbox{$q$}}
\newcommand{\slR}{\raise.15ex\hbox{$/$}\kern-.53em\hbox{$R$}}
\newcommand{\slQ}{\raise.15ex\hbox{$/$}\kern-.53em\hbox{$Q$}}
\newcommand{\slK}{\raise.15ex\hbox{$/$}\kern-.53em\hbox{$K$}}
\newcommand{\slk}{\raise.15ex\hbox{$/$}\kern-.53em\hbox{$k$}}
\newcommand{\slD}{\raise.15ex\hbox{$/$}\kern-.73em\hbox{$D$}}
\newcommand{\slC}{\raise.15ex\hbox{$/$}\kern-.53em\hbox{$C$}}
\newcommand{\slA}{\raise.15ex\hbox{$/$}\kern-.53em\hbox{$A$}}
\newcommand{\slSigma}{\raise.15ex\hbox{$/$}\kern-.53em\hbox{$\Sigma$}}
\newcommand{\slpartial}{\raise.15ex\hbox{$/$}\kern-.53em\hbox{$\partial$}}
\newcommand{\slcalP}{\raise.15ex\hbox{$/$}\kern-.63em\hbox{$\cal P$}}

\newcommand{\overbar}[1]{\mkern 1.5mu\overline{\mkern-1.5mu#1\mkern-1.5mu}\mkern 1.5mu}
\newcommand{\mbar}{\overbar{m}}

\def\p{{\boldsymbol p}}

\def\k{{\boldsymbol k}}

\def\x{{\boldsymbol x}}
\def\y{{\boldsymbol y}}

\def\r{{\boldsymbol r}}

\def\u{{\boldsymbol u}}

\def\s{{\boldsymbol s}}

\newcommand{\beq}{\begin{equation}}
\newcommand{\eeq}{\end{equation}}

\catcode`\@=11

\newcount\@tempcntc
\def\@citex[#1]#2{\if@filesw\immediate\write\@auxout{\string\citation{#2}}\fi
  \@tempcnta\z@\@tempcntb\m@ne\def\@citea{}\@cite{%
        \@for\@citeb:=#2\do%
    {\@ifundefined{b@\@citeb}%
        {\@citeo\@tempcntb\m@ne\@citea%
                \def\@citea{,\penalty\@m\ }{\bf ?}\@warning%
                {Citation `\@citeb' on page \thepage \space undefined}}%
        {\setbox\z@\hbox{\global\@tempcntc0\csname b@\@citeb\endcsname\relax}
     \ifnum\@tempcntc=\z@ \@citeo\@tempcntb\m@ne%
       \@citea\def\@citea{,\penalty\@m}%
       \hbox{\csname b@\@citeb\endcsname}%
     \else%
      \advance\@tempcntb\@ne%
      \ifnum\@tempcntb=\@tempcntc%
      \else\advance\@tempcntb\m@ne\@citeo%
      \@tempcnta\@tempcntc\@tempcntb\@tempcntc\fi\fi}}\@citeo}{#1}}%

\def\@citeo{\ifnum\@tempcnta>\@tempcntb\else\@citea
  \def\@citea{,\penalty\@m}%
  \ifnum\@tempcnta=\@tempcntb\the\@tempcnta\else
   {\advance\@tempcnta\@ne\ifnum\@tempcnta=\@tempcntb \else
\def\@citea{--}\fi
    \advance\@tempcnta\m@ne\the\@tempcnta\@citea\the\@tempcntb}\fi\fi}

\catcode`\@=12

\title{\bf Isotropization of a longitudinally expanding system of scalar fields in the 2PI formalism}

\author[a]{Fran\c cois Gelis,}
\author[a]{Sigtryggur Hauksson}
\affiliation[a]{Institut de Physique Th\'eorique\\
  CEA/Saclay, Universit\'e Paris-Saclay\\
  91191 Gif sur Yvette, France}

\emailAdd{francois.gelis@ipht.fr}
\emailAdd{sigtryggur.hauksson@ipht.fr}

\abstract{
Motivated by isotropization of QCD matter in the initial stages of heavy-ion collisions, 
we consider a system of scalar fields that undergoes a boost invariant longitudinal expansion. We use the framework of the two-particle irreducible (2PI) effective action, which is close to the underlying quantum field theory, and resum self-energy corrections up to three loops. The resulting 2PI equations of motion are expressed in terms of the Milne coordinates to account for longitudinal expansion. By solving numerically these equations of motion, we can extract the occupation density and the effective mass generated by in-medium interactions. At the largest values of the coupling considered in this study, we observe the onset of isotropization both in the occupation number and in the momentum dependence of the effective mass.
}

\begin{document}
\maketitle

\section{Introduction}
High energy heavy ion collisions are performed at the Large Hadron Collider and at the Relativistic Heavy Ion Collider in order to probe the properties of deconfined nuclear matter. A striking finding is that the system formed in these collisions expands as a relativistic fluid with extremely small viscosity to entropy ratio, and that this hydrodynamical expansion starts very shortly after the collision \cite{Gale:2013da,Romatschke:2017ejr}. This raises the question of how the far-from-equilibrium QCD medium formed in the initial collision evolves in time and quickly becomes amenable to a hydrodynamic description.
Answering this question requires simulating many-body QCD out of thermal equilibrium.

A widely used description of the early instants of a heavy ion collision is based on the Color Glass Condensate (CGC), in which the system is made of a strong color field produced by the color charges contained in the two projectiles \cite{Iancu:2000hn,Ferreiro:2001qy,Iancu:2003xm,Gelis:2010nm}. 
At zeroth order in coupling, this color field is purely classical and is characterized by a single dimensionful scale, the saturation momentum $Q_s$ \cite{McLerran:1993ni,McLerran:1993ka,Krasnitz:1998ns,Krasnitz:1999wc,Krasnitz:2001qu}. The associated stress-energy tensor does not evolve as predicted by hydrodynamics. Indeed, the longitudinal pressure is initially negative \cite{Lappi:2006fp,Fukushima:2011nq}; by a time $\tau\sim Q_s$, it becomes slightly positive but remains much lower than the transverse pressure for all subsequent times (the ratio $P_{L}/P_{_T}$ decreases like $\tau^{-2}$, consistent with free streaming). In contrast, viscous hydrodynamics displays isotropization of the stress-energy tensor when started with an anisotropic initial condition.

Beyond zeroth order, some loop corrections to the CGC description exhibit instabilities (i.e., an exponential growth in time) \cite{Romatschke:2005pm,Romatschke:2006nk} and need to be resummed for a consistent calculation. Such a resummation can be performed via the classical statistical approximation \cite{Dusling:2010rm,Gelis:2010nm}, which amounts to letting the initial condition at $\tau=0^{+}$ for the CGC classical field fluctuate with Gaussian fluctuations\footnote{For an appropriately chosen distribution of these Gaussian fluctuations, the classical statistical approximation reproduces exactly the leading and next-to-leading orders \cite{Epelbaum:2013waa,Epelbaum:2013ekf}. In addition, it resums an infinite subset of higher order corrections.}, and to have a classical evolution afterwards\footnote{This scheme, in which the evolution remains classical, can nevertheless give the exact one-loop correction. Indeed, it is a standard result of quantum mechanics that the order $\hbar$ correction comes entirely from the wave-function of the initial state, while corrections of order $\hbar^2$ and higher come both from the initial wave-funtion and from the time evolution. \cite{Groenewold:1946kp,Moyal:1949sk}}. However, this approach has some shortcomings. If the Gaussian fluctuations do not include zero-point vacuum fluctuations, then the scheme leads to ultraviolet finite results but lacks some terms essential for isotropization \cite{Berges:2013eia,Berges:2013fga,Berges:2015ixa,Epelbaum:2015vxa}. If vacuum fluctuations are included, then the longitudinal to transverse pressure ratio displays a trend towards isotropization \cite{Epelbaum:2013ekf}, but the classical statistical approximation becomes questionable because it lacks a continuum limit \cite{Epelbaum:2014yja}. 

Departing from the CGC treatment, kinetic theory has also been employed to study the relaxation of an out-of-equilibrium system in the context of heavy ion collisions \cite{Xu:2004mz,Arnold:2002zm,Molnar:2000jh,Martinez:2010sc,Florkowski:2013lya,Bazow:2013ifa,Florkowski:2013lza,Denicol:2014xca,Kurkela:2015qoa,Epelbaum:2015vxa,Kurkela:2018vqr,Kurkela:2018wud,Kamata:2020mka}. In QCD the medium constituents are then quark and gluon quasiparticles that interact occasionnally through 2-to-2 scattering or medium-induced radiation. This approach can describe the onset of isotropization. However, it cannot capture the very first instances of collisions when the occupation density of gluons is extremely high. Furthermore, even though Debye screening of the scatterings may be taken into account in kinetic theory, it cannot account for the dynamical in-medium modification of the quasiparticle masses, which may make it unreliable at describing the soft particles.  Yet another approach used to study this out-of-equilibrium evolution is the gauge/gravity duality valid at strong coupling, see for instance \cite{Janik:2005zt,Janik:2006gp,Heller:2007qt,Beuf:2009cx,Heller:2011ju,Casalderrey-Solana:2011dxg,Heller:2012km,Heller:2012je}.
Simulating the initial stages of heavy-ion collisions is not only interesting in itself but also gives information about experimental probes at early times. A particularly relevant observable is 
the transport coefficient $\widehat{q}$ that controls momentum broadening and energy loss of jet particles \cite{Boguslavski:2023waw, Boguslavski:2023jvg,Hauksson:2021okc,Ipp:2020mjc, Avramescu:2023qvv,Carrington:2022bnv, Carrington:2021dvw}.

The two-particle irreducible (2PI) effective action \cite{Berges:2004yj} has the potential to resolve many of the open questions about the early stages of heavy-ion collisions. In particular, it is a renormalizable scheme that can include all in-medium effects, and would therefore avoid the shortcomings of the classical statistical approximation and of kinetic theory. The 2PI effective action, originally introduced by Luttinger and Ward in solid state physics \cite{Luttinger:1960ua}, is a formulation of quantum field theory where \(1\)- and \(2\)-point functions of fields are evolved in time. The equations of motion come from resumming a subset of diagrams of the full quantum field theory to all orders. This is therefore closer to the underlying full theory. Moreover, truncations in this scheme have the virtue of preserving the conservations laws of the system \cite{Baym:1961zz}. Since its inception, the formalism of the 2PI effective action has been employed in a number of areas, such as cosmology \cite{Tranberg:2008ae}, systems of cold atoms \cite{Schmiedmayer:2013xsa,Berges:2015kfa}, the study  of a system of fields in a fixed volume \cite{Arrizabalaga:2004iw,Arrizabalaga:2005tf,Alford:2004jj,Berges:2004hn,Berges:2010nk,Tsutsui:2017uzd}. It was also employed for a system with the longitudinally expanding geometry encountered in high energy collisions in \cite{Hatta:2012gq}, albeit with limited grid sizes.


In principle, the 2PI effective action can be used to describe the whole of the initial stages of heavy-ion collisions in a single unified framework. This is because the effective action encompasses 
both kinetic theory and classical field theory, the main theoretical tools used to describe the initial stages, in different limits. For instance, kinetic theory is obtained from the 2PI effective action by applying a quasiparticle ansatz, so that all particles are assumed to be on-shell, and by applying an expansion in spacetime gradients, meaning that the medium is assumed to evolve slowly \cite{KadanoffBaymBook}.  The effective action furthermore allows for very high occupancy as in classical field theory. It therefore offers a smooth interpolation between classical field theory and kinetic theory, while going beyond both of these descriptions in allowing for off-shell excitations, abrupt changes in time and high densities.
Another advantage of the 2PI effective action is that it describes the medium dynamically across all energy scales, including soft modes that are not dynamical in kinetic theory simulations.
It can furthermore describe a system even if it is not made of well defined
quasi-particles.

Despite these advantages, the 2PI effective action has a major drawback. It remains unclear how to apply it  to non-Abelian gauge theories such as QCD in a gauge-invariant and fully renormalized manner \cite{Hatta:2011ky}. Therefore applications have been restricted to toy models such as scalar field theory. 
Such toy-model calculations are nevertheless very useful. Despite its simplicity, a longitudinally expanding medium of scalars shares many characteristics with the QCD medium in heavy-ion collisions, such as isotropization, coupling of soft and hard modes and the potential presence of instabilities. The major difference is that chemical equilibration is much slower than in QCD. Additionally, kinetic theory and classical-statistical calculations using scalar field theory can be benchmarked by 2PI calculations. Thus one can test the quality of approximations made in heavy-ion collisions in a simpler setup. Most importantly, the 2PI effective action makes it possible to simulate quantum fields from nearly first principles, albeit for a simpler theory than QCD.

The 2PI effective action for scalar fields has found numerous applications, most of which have been in a fixed box. Early work focused on thermalization in 1+1 D systems \cite{Berges:2001fi,Berges:2000ur,Aarts:2001qa, Aarts:2001yn} both in a loop expansion and in an large \(N\) expansion \cite{Aarts:2002dj}. Since then, there have been extensions to 3+1 D \cite{Juchem:2003bi}, and additionally the inclusion of fermions \cite{Berges:2002wr} and broken phases \cite{Arrizabalaga:2005tf}. The 2PI effective action has furthermore been used to analyze isotropization in a static box \cite{Berges:2005ai}, the presence of non-thermal fixed points beyond weak coupling \cite{Berges:2008wm,Berges:2016nru,Preis:2022uqs}, decoherence of classical states \cite{Kovtun:2020udn}, the formation of Bose-Einstein condensates \cite{Tsutsui:2017uzd} and parametric resonances \cite{Berges:2002cz,Arrizabalaga:2004iw}.

Heavy-ion collisions have a more complicated geometry than applications in a fixed box. Due to longitudinal expansion of the medium, the longitudinal momentum \(p_z\) drops rapidly. Furthermore, the system produced in these collisions is approximately invariant under boosts in the longitudinal direction. It is therefore more convenient to use proper time \(\tau\) and rapidity \(\eta\), as well as the momentum variable corresponding to rapidity which we call \(\nu \approx  p_z/\tau\). In an interacting medium the typical value of \(\nu\) grows with time. Therefore numerical calculations require a lattice which has a large extent in \(\nu\) and which is also fine enough to resolve early-time behaviour. This demands considerable computational resources. For this reason, and due to other technical challenges that will be detailed in the main text, the 2PI effective action has found  little application in the context of heavy-ion collisions, apart from a pioneering proof-of-concept work in \cite{Hatta:2012gq}. The slightly simpler case of an isotropic expansion as in an expanding universe was treated in  \cite{Tranberg:2008ae, Aarts:2007ye, Rajantie:2006gy} using the 2PI effective action. In this paper we develop an independent formulation of the 2PI effective action for a longitudinally expanding medium. We focus on the competition between interactions which tend to isotropize the medium and expansion which drives it away from isotropy. We also describe results that are unique to the 2PI framework such as off-shell excitations and dynamical soft modes.

This paper is organized as follows. In Sec. \ref{sec:2PI_formalism} we discuss the 2PI formalism and derive its equations of motion using a truncation in the number of loops. In Sec. \ref{sec:Milne_coordinates} we discuss how the 2PI equations of motion should be treated for a longitudinally expanding medium. We furthermore discuss how we renormalize those equations. Sec. \ref{sec:num_implem} presents how these equations can be implemented numerically in an efficient manner and Sec. \ref{sec:Results} contains the main results of our work. These results include isotropization in the occupation density of quasiparticles, the momentum dependence of the particles' thermal mass and the importance of number-changing processes. Finally, we conclude in Sec. \ref{sec:concl} and discuss future directions. Technical material on the Fourier transform and the form of bare propagators in our coordinate system, as well as the quasi-particle approximation is relegated to appendices.

\section{Two-particle irreducible formalism}
\label{sec:2PI_formalism}
We consider in this paper the theory of a real scalar field, with a
quartic interaction, in four spacetime dimensions. The Lagrangian
density reads
\begin{align}
  {\cal L}\equiv \frac{1}{2}(\partial_\mu\phi)(\partial^\mu\phi)-\frac{m^2}{2}\phi^2 -\frac{g^2}{4!}\phi^4.
  \label{eq:lagrangian}
\end{align}
(By analogy with QCD, we denote by $g^2$ the coupling constant of the
quartic interaction. This notation is natural when discussing
quantities such as thermal masses, that are of order $gQ$, where $Q$ is
the typical particle momentum in the system.)

\subsection{2PI quantum effective action}
The starting point to derive the Kadanoff-Baym equations is the 2PI
quantum effective action,
\begin{align}
\Gamma[\varphi,G]
=
S[\varphi]-\frac{i}{2}{\rm tr}\,\big(\log G\big)
+\frac{i}{2}{\rm tr}\,\big(\big(G_0^{-1}-G^{-1}\big)G\big)+\Phi[\varphi,G],
\label{eq:Gamma}
\end{align}
where $S[\varphi]$ is the classical action. The variables $\varphi$
and $G$ are, respectively, a 1-point function (the expectation value of
the field operator) and a 2-point function (the expectation value of
the propagator, defined on the two branches of the Schwinger-Keldysh
contour). $G_0^{-1}$ is defined as the second derivative of the
classical action with respect to the field $\varphi$ (up to a factor
$i$, such that $G_0$ is normalized in the same fashion as the
propagator $G$). In Cartesian coordinates, it reads
\begin{equation}
G_0^{-1}\equiv i\left(\square_x+m^2+\tfrac{g^2}{2}\varphi^2(x)\right)\delta(x-y).
\end{equation}
Thus, this object obeys
\begin{align}
  \left(\square_x+m^2+\tfrac{g^2}{2}\varphi^2(x)\right)G_0(x,y)=-i\delta(x-y).
\end{align}
In eq.~(\ref{eq:Gamma}), $\Phi[\varphi,G]$ is a functional of
$\varphi$ and $G$ whose diagrammatic representation is the sum of all
vacuum graphs that are two-particle-irreducible, whose lines are made
of the propagator $G$ and with insertions of the field $\varphi$.

\subsection{Equations of motion}
The equations of motion in this framework are
\begin{align}
  \frac{\delta \Gamma}{\delta \varphi(x)}=0,\quad
  \frac{\delta \Gamma}{\delta G(x,y)}=0.
\label{eq:eom1}
\end{align}
More explicitly, these equations read
\begin{align}
  &(\square_x\!+\!m^2) \varphi(x) + \tfrac{g^2}{6}\varphi^3(x) +\tfrac{g^2}{2}G(x,x)\varphi(x)=\frac{\delta\Phi}{\delta\varphi(x)},\nonumber\\
  &(\square_x\!+\!m^2)G(x,y)+\tfrac{g^2}{2}\varphi^2(x)G(x,y)=-i\delta(x-y)
  +\int d^4z\,{\frac{\delta \Phi}{\delta G(x,z)}}\,G(z,y).
\label{eq:eom2}
\end{align}
These equations are exact (provided one uses the exact
$\Phi[\varphi,G]$), and their solution fully determines the evolution
of the system given some initial conditions.

\subsection{Truncation of the effective action at order $g^4$}
In order to make a practical use of these equations of motion, it is
necessary to have an explicit form for the 2PI functional
$\Phi[\varphi,G]$. The simplest way to achieve this is to truncate its
diagrammatic expansion at some fixed order\footnote{In theories with
$N$ fields, it is possible to organize the diagrammatic expansion in
powers of $1/N$. When $N$ is large, it is then possible to resum all the
leading contributions in this expansion.}.  In the ``sunset
approximation'', that we use in this paper, the $\Phi$ functional is
truncated at order $g^4$: \setbox1\hbox to
40mm{\includegraphics[width=40mm]{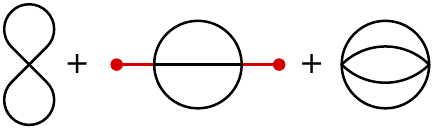}}
\begin{equation}
i\Phi[\varphi,G]\empile{=}\over{\mbox{\scriptsize order $g^4$}}\;\;
\raise -5mm\box1\;.
\label{eq:sunset_app}
\end{equation}
(The red dots indicate insertions of the field expectation
value $\varphi$ and the solid black lines are the propagator $G$. From
left to right, the symmetry factors of these graphs are respectively
$1/8$, $1/12$ and $1/48$.) 
This approximation is the lowest order truncation that includes the
effect of $2\to 2$ scatterings. Indeed, if one performs a gradient
expansion and a quasi-particle approximation, the 2PI equations of
motion with the truncation (\ref{eq:sunset_app}) lead to a Boltzmann
equation with elastic two-body collisions.

In terms of the propagator $G$ and the field $\varphi$, the three
graphs in the r.h.s of eq.~(\ref{eq:sunset_app}) read
\setbox1\hbox to 7mm{\includegraphics[width=7mm]{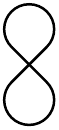}}
\setbox2\hbox to 21mm{\includegraphics[width=21mm]{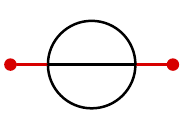}}
\setbox3\hbox to 13mm{\includegraphics[width=13mm]{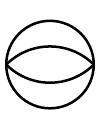}}
\begin{eqnarray}
\raise -7.0mm\box1 &=&-i\tfrac{g^2}{8}\int_{{\cal C}_{t_{\rm init}}} d^4z\; G^2(z,z),\nonumber\\
\raise -6.5mm\box2 &=&\tfrac{(-ig^2)^2}{12}\int_{{\cal C}_{t_{\rm init}}}d^4z d^4z'\; \varphi(z)\varphi(z')G^3(z,z'),\nonumber\\
\raise -8mm\box3 &=&\tfrac{(-ig^2)^2}{48}\int_{{\cal C}_{t_{\rm init}}}d^4z d^4z'\; G^4(z,z'),
\end{eqnarray}
where ${\cal C}_{t_{\rm init}}$ is the closed time path starting at some
initial time $t_{\rm init}$, extending to $+\infty$, and returning to $t_{\rm init}$.

 From this truncated 2PI
functional, it is easy to derive the expression of the right hand
sides of the equations of motion. At this order of truncation, the derivatives
$\delta\Phi/\delta\varphi(x)$ and $\delta\Phi/\delta G(x,y)$ that
enter in the r.h.s of the equations of motion are given by
\begin{align}
  \frac{\delta\Phi}{\delta\varphi(x)}
  & =
  i\tfrac{g^4}{6}\int_{{\cal C}_{t_{\rm init}}}d^4z\; G^3(x,z)\varphi(z),\nonumber\\
  \frac{\delta\Phi}{\delta G(x,y)}
  & = -\tfrac{g^2}{2}\delta(x-y)G(x,x)
  +i\tfrac{g^4}{2} \varphi(x)\varphi(y)G^2(x,y)
  +i\tfrac{g^4}{6} G^3(x,y).
\end{align}

\subsection{Statistical and spectral functions}
In the equations of motion (\ref{eq:eom2}), $G(x,y)$ is the path
ordered propagator defined on the Schwinger-Keldysh time
contour ${\cal C}_{t_{\rm init}}$. For a practical numerical resolution, it is
necessary to rewrite these equations so that integrals over
the Schwinger-Keldysh closed time contour are explicitly rewritten as
a pair of integrals over the ordinary real line.

First, note that the expectation value of the field is the same on the
upper and lower branches, i.e.,
\begin{align}
\varphi_+(x)=\varphi_-(x)=\varphi(x).
\end{align}
($\varphi_\pm(x)$ denotes the expectation value of the field operator with
$x^0$ on the upper and lower branches of the contour, respectively.)

The situation is more complicated for the propagator $G$. Firstly,
recall the following identity,
\begin{align}
  G_{++}(x,y)+G_{--}(x,y)=G_{+-}(x,y)+G_{-+}(x,y),
\end{align}
that relates the 2-point functions with the four possible assignments
for their time arguments. Instead of using the components $G_{\pm
  \pm}$ of the propagator, it is more convenient to use the
retarded/advanced basis in which the propagator is decomposed into
into a statistical function $F(x,y)$ and a spectral function
$\rho(x,y)$. Their relationship with the original Schwinger-Keldysh
propagators are given by
\begin{align}
  &F(x,y)\equiv \tfrac{1}{2}\big(G_{-+}(x,y)+G_{+-}(x,y)\big),\nonumber\\
  &\rho(x,y)\equiv -i \big(G_{-+}(x,y)-G_{+-}(x,y)\big).
  \label{eqs:Frho}
\end{align}
Inversely, we have:
\begin{align}
  &G_{-+}(x,y)=F(x,y)+\tfrac{i}{2}\rho(x,y),\nonumber\\
  &G_{+-}(x,y)=F(x,y)-\tfrac{i}{2}\rho(x,y),\nonumber\\
  &G_{++}(x,y)=F(x,y)+\tfrac{i}{2}{\rm sign}\,(x^0-y^0)\,\rho(x,y),\nonumber\\
  &G_{--}(x,y)=F(x,y)-\tfrac{i}{2}{\rm sign}\,(x^0-y^0)\,\rho(x,y).
\end{align}
Note that $F(x,y)$ is symmetric under the exchange
$x\longleftrightarrow y$, while $\rho(x,y)$ is
antisymmetric. Moreover, since
$G_{+-}(x,y)\equiv\big<\phi(y)\phi(x)\big>$ is the complex conjugate
of $G_{-+}(x,y)\equiv\big<\phi(x)\phi(y)\big>$, we see easily that
$F(x,y)$ and $\rho(x,y)$ are both real valued (thanks to the
prefactor $i$ included in the definition of $\rho$).

\subsection{Equations of motion in terms of $\varphi,F,\rho$}
After explicitly separating the two branches of the Schwinger-Keldysh
time contour, and systematically trading the propagators $G_{\pm\pm}$
in favor of $F,\rho$, the equations of motion take the following form:
\begin{align}
\Big[\square_x+m^2 + M_0^2(x)\Big]\varphi(x)+\tfrac{g^2}{6}\varphi^3(x)
=
\int_{t_{\rm init}}^{x^0}d^4z\; \Pi_\rho(x,z)\,\varphi(z),
\label{eq:eom-phi}
\end{align}
\begin{align}
\Big[\square_x+m^2 + M_0^2(x)+\tfrac{g^2}{2}\varphi^2(x)\Big]\,F(x,y)
&=
\int_{t_{\rm init}}^{x^0}d^4z\;\Sigma_\rho(x,z)\,F(z,y)
\nonumber\\
&
+
\int_{t_{\rm init}}^{y^0}d^4z\;\Sigma_{_F}(x,z)\,\rho(z,y),
\label{eq:eom-F}
\end{align}
and
\begin{align}
\Big[\square_x+m^2 + M_0^2(x)+\tfrac{g^2}{2}\varphi^2(x)\Big]\,\rho(x,y)
=
\int_{y^0}^{x^0}d^4z\;\Sigma_\rho(x,z)\,\rho(z,y).
\label{eq:eom-rho}
\end{align}
In these equations, $M_0^2(x)$ is a mass contribution coming from the tadpole,
\begin{align}
  M_0^2(x)\equiv \frac{g^2}{2}F(x,x),
\end{align}
and the self-energies $\Pi_\rho$, $\Sigma_{_F}$ and $\Sigma_\rho$ are defined as
\begin{equation}
  \Sigma_{_F}(x,y)
  \equiv
  \frac{g^4}{2}\varphi(x)\varphi(y)
  \Big[F^2(x,y)\!-\!\frac{1}{4}\rho^2(x,y)\Big]
  +
  \frac{g^4}{6}F(x,y)\Big[F^2(x,y)-\frac{3}{4}\rho^2(x,y)\Big],
\end{equation}
\begin{equation}
  \Sigma_{\rho}(x,y)
  \equiv
  -g^4\varphi(x)\varphi(y)F(x,y)\rho(x,y)
  -\frac{g^4}{6}
  \rho(x,y)\Big[3F^2(x,y)-\frac{1}{4}\rho^2(x,y)\Big],
\end{equation}
and
\begin{equation}
  \Pi_\rho(x,y)
  \equiv
  -\frac{g^4}{6}
  \rho(x,y)\Big[3F^2(x,y)-\frac{1}{4}\rho^2(x,y)\Big].
\end{equation}
Thanks to the symmetries of $F$ and $\rho$, $\Sigma_{_F}(x,y)$ is also
real and symmetric, while $\Sigma_\rho(x,y)$ and $\Pi_\rho(x,y)$ are
real and antisymmetric. This ensures that $\varphi$, $F$ and $\rho$
remain real throughout their evolution.

\section{Formulation in Milne coordinates}
\label{sec:Milne_coordinates}
\subsection{Longitudinal boost invariance}
The equations (\ref{eq:eom-phi}), (\ref{eq:eom-F}) and
(\ref{eq:eom-rho}) are completely general, except for the truncation
at order $g^4$ of the 2PI effective action (which gives their explicit
form to the self-energies). In this paper, we wish to describe the
evolution of a system whose geometry mimics that of the system
produced in a heavy ion collision at high energy. In particular, we
assume that it is invariant under boosts along the $x^3$ direction
(the collision axis). This boost invariance is more easily expressed
if we use the spatial rapidity $\eta\equiv
\log((x^0+x^3)/(x^0-x^3))/2$ instead of the Cartesian coordinate
$x^3$, and the proper time $\tau\equiv\sqrt{(x^0)^2-(x^3)^2}$ instead
of $x^0$. Indeed, $\tau$ is invariant under these boosts, while $\eta$
is simply translated by a constant shift. Thus, the longitudinal boost
invariance implies that $\varphi$ is independent of rapidity, while
the 2-point functions $F$ and $\rho$ depend only on the rapidity
difference. Furthermore, for a collision of two identical projectiles, the system is mirror-symmetric about the
plane $x^3=0$ and these two-point functions are even in the rapidity
difference, i.e., depend only on $|\eta_x-\eta_y|$:
\begin{align}
  \varphi(x)&=\varphi(\tau_x,\x_\perp), \nonumber\\
  F(x,y)&=F(\tau_x,\x_\perp,\tau_y,\y_\perp,|\eta_x-\eta_y|),\nonumber\\
  \rho(x,y)&=\rho(\tau_x,\x_\perp,\tau_y,\y_\perp,|\eta_x-\eta_y|).
\end{align}
(We denote $\x_\perp,\y_\perp,\cdots$ the 2-dimensional vector
representing the coordinates in the plane orthogonal to the $x^3$
axis.) In order to rewrite the equations of motion in this system of
coordinates, we need
\begin{align}
  \square = \partial_\tau^2+\tfrac{1}{\tau}\partial_\tau - {\boldsymbol\partial}_\perp^2 -\tfrac{1}{\tau^2}\partial_\eta^2,\quad d^4x=\tau d\tau \,d\eta\, d^2\x_\perp.
\end{align}

\subsection{Translation and rotation invariance in the transverse plane}
We further assume that the system is invariant under translations and
rotations in the transverse plane. This assumption implies that the
field expectation value depends only on the proper time:
\begin{align}
\varphi(x)=\varphi(\tau_x).
\end{align}
For the two-point functions $F(x,y)$ and $\rho(x,y)$, the
$\x_\perp,\y_\perp$ dependence arises only through the relative transverse distance
$|\x_\perp-\y_\perp|$:
\begin{align}
  F(x,y)&=F(\tau_x,\tau_y,|\x_\perp-\y_\perp|,|\eta_x-\eta_y|),\nonumber\\
  \rho(x,y)&=\rho(\tau_x,\tau_y,|\x_\perp-\y_\perp|,|\eta_x-\eta_y|).
\end{align}

\subsection{Spatial Fourier transform}
Thanks to the translation invariance in the transverse plane and along
the rapidity direction, it will prove convenient to introduce the
following Fourier transforms:
\begin{align}
  F(\tau_x,\tau_y,\p_\perp,\nu)&\equiv
  \int d^2\s_\perp\,d\eta \;
  e^{i\p_\perp\cdot\s_\perp}\,
  e^{i\nu\eta}\;
  F(\tau_x,\tau_y,|\s_\perp|,|\eta|),\nonumber\\
  \rho(\tau_x,\tau_y,\p_\perp,\nu)&\equiv
  \int d^2\s_\perp\,d\eta \;
  e^{i\p_\perp\cdot\s_\perp}\,
  e^{i\nu\eta}\;
  \rho(\tau_x,\tau_y,|\s_\perp|,|\eta|),
\end{align}
(We use the same symbols for the Fourier transformed propagators, as
the context is always sufficient to know whether an equation contains
propagators in coordinate or in momentum space.) Since the propagators $F$ and $\rho$ are both even in $\s_\perp$ and
in $\eta$, their Fourier transforms are real valued.  Note that since
the system of interest is out-of-equilibrium, there is no invariance
under translations in time, and these Fourier transforms depend
separately on the two times $\tau_x$ and $\tau_y$, with $F$
(resp. $\rho$) symmetric (resp. antisymmetric) under the exchange of
$\tau_x$ and $\tau_y$.

In order to gain some intuition for the variable $\nu$, the Fourier
conjugate to the rapidity difference, it is useful to have in mind the
following back-of-the-envelope correspondence between $\nu$ and the
third component of the spatial momentum, \(p_z\). Let us start from the
relationship between the derivatives with respect to the coordinates
$(x^0,x^3)$ and $(\tau,\eta)$:
\begin{align}
  \frac{\partial}{\partial x^3}
  &=
  -\sinh(\eta)\frac{\partial}{\partial\tau}
  +
  \tau^{-1}\cosh(\eta)\frac{\partial}{\partial\eta},\nonumber\\
  \frac{\partial}{\partial x^0}
  &=
  \cosh(\eta)\frac{\partial}{\partial\tau}
  -
  \tau^{-1}\sinh(\eta)\frac{\partial}{\partial\eta}.
\end{align}
In the vicinity of $\eta=0$, these relations simplify into
\begin{align}
  \frac{\partial}{\partial x^3}
 \approx
  \tau^{-1}\frac{\partial}{\partial\eta},\quad
  \frac{\partial}{\partial x^0}
  \approx
  \frac{\partial}{\partial\tau}.
  \label{eq:der-rel}
\end{align}
Then, by identifying the spatial derivatives with the corresponding momenta, we obtain
\begin{align}
\nu \approx \tau\, p_z.
\end{align}
Note that, if the system becomes isotropic, then the typical momenta would decrease with proper time roughly as $p_\perp,p_z\sim \tau^{-1/3}$. This would imply that the typical $\nu$ grows as $\nu\sim \tau^{2/3}$. Consequently, for a discretization with a maximal value $\nu_{\rm max}$, there is always a time beyond which the longitudinal momenta of the particles of an isotropic system do not fit anymore on the grid. Studying numerically the approach to isotropy therefore requires a discretization with as large a $\nu_{\rm max}$ as possible. This limitation does not exist for the study of a system undergoing free-streaming. Indeed, for free-streaming, the typical $p_z$ decreases as $p_z\sim \tau^{-1}$ and the typical $\nu$ is constant.

\subsection{Equations of motion with the variables $\tau,p_\perp,\nu$}
In a non-interacting theory, the dynamical evolution of the system is
separable in the variables $p_\perp,\nu$, because there is no term in
the free equations of motion that can mix the Fourier modes. This
choice of variables is therefore very appropriate to follow the
evolution of a  system where the
interactions provide a moderate correction to the free streaming
evolution. In this system of coordinates, the equations of motion read
\begin{align}
\Big[\partial_\tau^2+\tfrac{1}{\tau}\partial_\tau+m^2 + M_0^2(\tau)\Big]\varphi(\tau)+\tfrac{g^2}{6}\varphi^3(\tau)
=
\int_{\tau_{\rm init}}^{\tau}d\tau'' \tau''\; \Pi_\rho(\tau,\tau'',p_\perp\!\!=\!\nu\!=\!0)\,\varphi(\tau''),
\label{eq:eom-phi-p}
\end{align}
\begin{align}
&\Big[\partial_\tau^2+\tfrac{1}{\tau}\partial_\tau+m^2 +M_0^2(\tau)+\tfrac{g^2}{2}\varphi^2(\tau)+p_\perp^2+\tfrac{\nu^2}{\tau^2}\Big]\,F(\tau,\tau',\p_\perp,\nu)\nonumber\\
&\qquad=
\int_{\tau_{\rm init}}^{\tau}d\tau''\tau''\;\Sigma_\rho(\tau,\tau'',p_\perp,\nu)\,F(\tau'',\tau',p_\perp,\nu)
\nonumber\\
&
\qquad+
\int_{\tau_{\rm init}}^{\tau}d\tau''\tau''\;\Sigma_{_F}(\tau,\tau'',p_\perp,\nu)\,\rho(\tau'',\tau',p_\perp,\nu),
\label{eq:eom-F-p}
\end{align}
and
\begin{align}
&\Big[\partial_\tau^2+\tfrac{1}{\tau}\partial_\tau+m^2+M_0^2(\tau)+\tfrac{g^2}{2}\varphi^2(\tau)+p_\perp^2+\tfrac{\nu^2}{\tau^2}\Big]\,\rho(\tau,\tau',\p_\perp,\nu)\nonumber\\
&\qquad=
\int_{\tau'}^{\tau}d\tau''\tau''\;\Sigma_\rho(\tau,\tau'',p_\perp,\nu)\,\rho(\tau'',\tau',p_\perp,\nu).
\label{eq:eom-rho-p}
\end{align}
The only non-locality in momentum space is now hidden in the self-energies $\Pi_\rho,\Sigma_\rho,\Sigma_{_F}$, that contain convolutions over the momenta.
In these equations, the effective mass $M_0^2(\tau)$ is defined as
\begin{align}
\label{Eq:M2_unrenorm}
  M_0^2(\tau)\equiv \tfrac{g^2}{2}\int \frac{dp_\perp p_\perp\,d\nu}{(2\pi)^2}\,F(\tau,\tau,p_\perp,\nu).
\end{align}

\subsection{Renormalization}
\label{sec:renorm}

\begin{figure}
\centering
\begin{subfigure}{0.075\textwidth}
    \includegraphics[width=\textwidth]{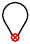}
\end{subfigure}
\hskip 10mm
\begin{subfigure}{0.075\textwidth}
    \includegraphics[width=\textwidth]{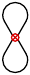}
\end{subfigure}
\hskip 10mm
\begin{subfigure}{0.075\textwidth}
    \includegraphics[width=\textwidth]{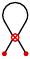}
\end{subfigure}
\hskip 10mm
\begin{subfigure}{0.075\textwidth}
    \includegraphics[width=\textwidth]{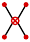}
\end{subfigure}
\caption{Additional contributions to \(\Phi\) that include counterterms.}
\label{fig:Ren_diagrams}
\end{figure}

The equations of motion in Eqs. \eqref{eq:eom-phi-p}, \eqref{eq:eom-F-p}, \eqref{eq:eom-rho-p}  contain ultraviolet divergences both in the memory integrals and in the tadpole \(M_0^2\). Discretizing these equations on a lattice renders them finite, however a rigorous implementation requires removing the ultraviolet divergences through renormalization.  
In the 2PI framework, this is paramount to removing all divergences from diagrams in \(\Phi\). At first sight this might seem difficult  as each diagram in \(\Phi\) describes a partial resummation to all orders of diagrams in perturbation theory and these perturbative diagrams might not have a simple structure in the UV. However, it has been shown that introducing a  handful of counterterms gives new resummed diagrams in \(\Phi\), shown in Fig. \ref{fig:Ren_diagrams}, 
which exactly cancel all divergences \cite{Blaizot:2003an}. This procedure gives equations of motion that are finite and independent of the details of regularization and which respect all symmetries of the theory \cite{Berges:2005hc}. This procedure has also been shown to work for QED \cite{Reinosa:2009tc}.

The counterterms needed in Fig. \ref{fig:Ren_diagrams} are fixed by renormalization conditions which define the strength of coupling \(g^2_{\mathrm{ren}}\), the squared mass  \(m^2_{\mathrm{ren}}\) and the field strength \(Z_{\mathrm{ren}}\) at a given momentum scale. Finding the counterterms then amounts to solving non-linear equations for vacuum propagators that come from the 2PI effective action \cite{Berges:2004hn,Borsanyi:2008ar}. These equations are only well-defined given a regularization scheme which must of course match the one used in the 2PI simulations.

In 2PI simulations in a fixed box the regularization scheme arises naturally: it is given by the cutoff \(\Lambda\) of the lattice in momentum space. In our setup with longitudinal expansion the regularization scheme is more complicated because there are two independent momentum cutoffs \(\Lambda_{\perp}\) and \(\Lambda_z\) for \(p_{\perp}\) and \(p_z = \nu/\tau\) and \(\Lambda_z\) changes with time. Specifically, the cutoff in the longitudinal direction is \(\Lambda_z(\tau) = \nu_{\mathrm{max}}/\tau\), where \(\nu_{\mathrm{max}}\) is the maximum value of \(\nu\) on the lattice. Therefore, in Cartesian coordinates the size of the lattice in momentum space shrinks with time. Instead of being able to find counterterms once and for all as in the case of a fixed regularization, one needs to find counterterms at each point in time. We plan to study this further in future work.\footnote{Full renormalization in the 2PI framework also requires introducing n-point sources in the initial state \cite{Garny:2009ni}, see the discussion in Sec. \ref{sec:setup_calc}. This may be understood as follows: with renormalization conditions set at a fixed energy, the counterterms involve loop integrals that extend over an infinite range of time (in practice, at least over a few Compton wavelengths). When $\tau$ is close to the initial time $\tau_0$, there is not enough time to capture the correct value of the counterterms and the missing part of the counterterms must be provided by extra ``sources'' included in the definition of the initial state. The effect of these sources pregressively disappears as $\tau$ increases.}

For our current purposes we adopt a much simpler renormalization procedure which relies on having a relatively weak coupling. Firstly, at sufficiently weak coupling the complicated non-linear equations that define the counterterms in the 2PI framework reduce to the same equations as in perturbation theory. Thus one can use perturbative counterterms.  Secondly, we will only remove quadratic divergences in the tadpole and not logarithmic divergences in the tadpole, as well as other divergences in the memory integral. This amounts to replacing the effective mass in Eq. \eqref{Eq:M2_unrenorm} by 
\begin{align}
\label{Eq:M2_unrenorm1}
  M_0^2(\tau)\equiv \tfrac{g^2}{2}\int \frac{dp_\perp p_\perp\,d\nu}{(2\pi)^2}\, \left[ F(\tau,\tau,p_\perp,\nu) - F_0(\tau,\tau,p_\perp,\nu) \right].
\end{align}
where \(F_0\) is the bare vacuum propagator.
The validity of this minimal renormalization procedure is ultimately verified by seeing how sensitive physical quantities are to the value of cutoffs. 
Below we will study how our results depend on varying the lattice size and see that the effect is modest \footnote{We have estimated that, for a coupling $g^4=500$, neglecting the coupling counterterm $\delta g^2$ leads to a variation of the results by about $15\%$ when we double the ultraviolet cutoffs. At this value of the coupling, the field strength counterterm $\delta Z$ is less than $1\%$.}.

To summarize this discussion on renormalization, let us list here the missing ingredients for a full renormalization of the 2PI equations of motion:
\begin{itemize}
    \item Our approximate renormalization of the mass is only correct to order $g^2$, but higher order corrections to the mass counterterm are necessary to deal with the divergences arising from order $g^4$ skeleton diagrams in $\Phi$.
    \item The 2-loop self-energy has an ultraviolet divergence proportional to the square of the momentum, that requires a counterterm for the field renormalization.
    \item A renormalization of the coupling is also necessary.
\end{itemize}

\section{Numerical implementation}
\label{sec:num_implem}
\subsection{Memory requirements}
With the symmetries we have assumed above, a propagator such as
$F(\tau_x,\tau_y,\cdots)$ depends on a transverse momentum
$p_\perp$ and a variable $\nu$ (or their counterparts in coordinate
space). Assume that the modulus of transverse momentum is discretized
with $N_\perp$ values, and that the variable $\nu$ is discretized with
$N_\eta$ values. Therefore, a propagator at fixed times
$\tau_x,\tau_y$ is represented by $N_\perp N_\eta$ floating point
numbers. 

A very important feature of the right hand sides of the equations of
motion (\ref{eq:eom-phi}), (\ref{eq:eom-F}) and (\ref{eq:eom-rho}) is
that they contain ``memory integrals'', i.e., integrals that depend on
the past history of the system. This aspect is quite challenging for
numerical approaches because, if taken literally, it would imply that
memory needs grow quadratically with the elapsed time since the
initial time. This can be mitigated by noting that the system should
loose the memory of its past evolution after a few collisions. In
other words, the range of times that contribute significantly to these
memory integrals should not exceed a few mean free paths. Therefore,
instead of storing the entire past evolution of the system, we store a
sliding window of the last $N_\tau$ timesteps. The overall memory
footprint for the storage of the propagators is thus
\begin{align}
  {\rm Memory} \approx 2\,N_\perp \frac{N_\eta}{2} \frac{N_\tau^2}{2}\quad \mbox{floating point numbers}.
\end{align}
(The prefactor $2$ is for $F$ and $\rho$, the first factor $1/2$ is
due to the parity of the propagators under the exchange $\eta\to
-\eta$, while the second factor $1/2$ comes from exploiting the
symmetry (resp. antisymmetry) in time of $F$ (resp. $\rho$).) For
instance, with $N_\perp=512$, $N_\eta=1536$ and $N_\tau=300$, and with
simple precision (i.e., $4$ Bytes per floating point number), we
need $132$ Giga-Bytes of memory for storing the propagators. Some
extra memory is needed for storing the field $\varphi$ and its past
history, as well as for the self-energies. Since for these items the
memory only grows linearly with $N_\tau$, these extra needs are
small compared to the propagators.

Given this estimate, it also seems impossible (with presently
available computing resources) to relax our assumptions about the
spatial symmetries of the system. For instance, for a system which is
not invariant under rotations in the transverse plane, we would need
to keep separately the dependence on the momenta $p_x$ and
$p_y$. Roughly speaking, this would transform the factor $N_\perp$
into $N_\perp^2$ in the above estimate, raising the needs to $\sim 66$
Tera-Bytes in the above example.

\subsection{Discretization of $\eta$ and $\nu$}
The discretization of the rapidity axis is the most
straightforward. We choose a constant spacing $a_\eta$ between two
successive values of $\eta$, and the discrete values we
consider are
\begin{align}
  \eta_i \equiv i\,a_\eta,\quad (0\le i< N_\eta).
\end{align}
The values of a function $f(\eta)$ will be denoted $f_i\equiv
f(\eta_i)$ after this discretization. Its Fourier transform is defined
as\footnote{There are several prescriptions for the value of $\nu$
corresponding to the index $k$. For instance, one may use
$\nu_\k=(2/a_\eta)\sin\big(\pi k/ N_\eta\big)$ or $\nu_k=2\pi
k/(a_\eta N_\eta)$, which coincide at small $k$ (i.e., in the region
far below the ultraviolet cutoff set by the lattice spacing).}
\begin{align}
  \tilde{f}_k \equiv a_\eta\sum_{i=0}^{N_\eta-1} f_i \,e^{i2\pi\tfrac{ki}{N_\eta}}.
  \label{eq:DFT}
\end{align}
The inverse transform reads
\begin{align}
  f_i \equiv \frac{1}{N_\eta a_\eta} \sum_{k=0}^{N_\eta-1} \tilde{f}_k \,e^{-i2\pi\tfrac{ki}{N_\eta}}.
  \label{eq:DFTinv}
\end{align}
This discrete Fourier transform satisfies a discrete form of the convolution
theorem. Namely, if a function is the point-wise multiplication of two
functions in coordinate space, $h_i=f_i g_i$, their discrete Fourier
transforms are related by
\begin{align}
  \tilde{h}_k = \frac{1}{N_\eta a_\eta}\sum_{p=0}^{N_\eta-1} \tilde{f}_p \,\tilde{g}_{k-p}.
\end{align}
This property will prove useful when we compute the self-energies, in
order to replace convolutions in momentum space by pointwise products
of the corresponding objects in coordinate space.

Note also that a discrete function representable as in
eq.~(\ref{eq:DFTinv}) is in fact periodic:
$f_{i+N_\eta}=f_i$. Therefore, the range of indices $N_\eta/2 < i <
N_\eta$ can also be viewed as describing the values of the function at
negative values of $\eta$. For an even function, that depends only on
$|\eta|$, the discretization therefore obeys
\begin{align}
  f_i = f_{N_\eta-i},
\end{align}
and we can save a factor $2$ in the memory needs by storing only the
first half of the range, $0\le i\le N_\eta/2$.


\subsection{Discretization of $r_\perp$ and $p_\perp$}
The discretization of the variables $|\r_\perp|$ and $|\p_\perp|$ is
more delicate, because we would like to discretize the radial
coordinate in the 2-dimensional plane (both in coordinate space and in
momentum space), and simultaneously have a discrete Fourier transform
relating the two and satisfying the convolution theorem.  The reason
why this is non-trivial is that the two-dimensional Fourier transform
of an axially symmetric function is in fact a Hankel transform (i.e.,
a one-dimensional transform whose kernel is a Bessel function instead
of a sine or cosine):
\begin{align}
  \int d^2\r_\perp \; e^{i\p_\perp\cdot \r_\perp}\; f(|\r_\perp|)
  &=
  \int_0^{+\infty} dr\,r\int_0^{2\pi} d\theta\; e^{ipr\cos(\theta)}\; f(r)
  \nonumber\\
  &= 2\pi \int_0^{+\infty} dr \,r\; J_0(pr)\;f(r).
  \label{eq:hankel}
\end{align}
There exists a discretization that permits a natural extension of this
transformation, known as the ``Quasi Discrete Hankel Transform''
(QDHT) \cite{qdht1,qdht2}. The starting point is to choose two numbers $R$ and $P$, which
represent the ranges covered by the discretization in $r_\perp$ and
$p_\perp$, respectively. Then, the discrete values of $r_\perp$ and
$p_\perp$ are defined to be
\begin{align}
  r_i\equiv \frac{\alpha_i}{P},\quad p_k\equiv \frac{\alpha_k}{R},
\end{align}
where the $\alpha_i$'s ($i=1,\cdots,N_\perp$) are the successive
zeroes of the Bessel function $J_0$. A function $f(r_\perp)$ is
replaced by a discrete set of values $f_i$, and its Fourier transform
$\tilde{f}(p_\perp)$ by discrete values $\tilde{f}_k$. The QDHT is
defined by the following pair of transforms:
\begin{align}
  \tilde{f}_k&=\frac{4\pi}{P^2}\sum_{i=1}^{N_\perp} f_i\;\frac{J_0\big(\tfrac{\alpha_i\alpha_k}{PR}\big)}{J_1^2\big(\alpha_k\big)},\nonumber\\
        {f}_i&=\frac{1}{\pi R^2}\sum_{i=1}^{N_\perp} \tilde{f}_k\;\frac{J_0\big(\tfrac{\alpha_i\alpha_k}{PR}\big)}{J_1^2\big(\alpha_i\big)}.
        \label{eq:qdht}
\end{align}
The word ``quasi'' in the name of this discrete
transform refers to the fact that these two formulas are not exact
inverses of each other at finite $N_\perp$, but only in the limit
$N_\perp\to \infty$. In fact, they are exact mutual inverses if the
following matrix,
\begin{equation}
  T_{ik}\equiv
  \frac{2}{PR}\frac{J_0(\tfrac{\alpha_i\alpha_k}{PR})}{|J_1(\alpha_i)J_1(\alpha_k)|},
\end{equation}
is equal to its own inverse, i.e., if $T_{ik}T_{kj}=\delta_{ij}$. This
property is true when $N_\perp\to \infty$, but only approximate for a
finite $N_\perp$. However, even at finite $N_\perp$, there are special
values of $P,R$ (such that their product $PR$ is close to
$\alpha_{N_\perp+1}$, the first zero of $J_0$ not used to construct
the discrete radii\footnote{At large $N_\perp$, the optimal $PR$ is
approximately $\alpha_{N_\perp+1}\approx \pi N_\perp$, and the maximal
values of $r_i$ and $p_k$ are therefore close to $R$ and $P$,
respectively.}) for which the violation of this identity is extremely
small and practically negligible.  For an explanation of the origin of
this transform and a discussion of some of its properties, see
Appendix \ref{app:qdht}.


\subsection{Calculation of the self-energies}
The equations of motion (\ref{eq:eom-phi-p}), (\ref{eq:eom-F-p}) and
(\ref{eq:eom-rho-p}) are seemingly local in $p_\perp,\nu$. However,
they contain the self-energies $\Pi_\rho, \Sigma_{_F},\Sigma_\rho$
that, at order \(g^4\), are defined as products in coordinate space and are therefore convolutions in momentum space. 
In our numerical calculations we replace these convolutions in momentum space 
by a pair of Fourier transforms and a point-wise ordinary
product in coordinate space: 

\begin{center}
\begin{tikzcd}[column sep=huge]
{F,\rho(\tau,\tau',p_\perp,\nu)} \arrow[r, "\text{convolution}"] \arrow[d, "\text{Fourier trans.}" left]
& {\Pi_\rho,\Sigma_{\rho,_F}(\tau,\tau',p_\perp,\nu) }  \\
{F,\rho(\tau,\tau',r_\perp,\eta)} \arrow[r, "\text{point-wise prod.}"]
&  {\Pi_\rho,\Sigma_{\rho,_F}(\tau,\tau',r_\perp,\eta) } \arrow[u, "\text{Fourier trans.}" right]
\end{tikzcd}
\end{center}
Thus with this approach, at every timestep, we Fourier transform the
propagators $F,\rho(\tau,\tau',p_\perp,\nu)$ to coordinate space, we
obtain the self-energies in coordinate space by an ordinary product,
and we Fourier transform the self-energies to momentum space.

\subsection{Evolution by one timestep}
The equations of motion for the propagators $F$ and $\rho$ have the structure of a Bessel equation modified by non-linear terms,
\begin{equation}
\big(\partial_\tau^2+\tau^{-1}\partial_\tau +(\underbrace{p^2_\perp+m^2}_{m_\perp^2}+\tfrac{\nu^2}{\tau^2})\big)\,X
={\rm RHS}(\tau),
\label{eq:ode}
\end{equation}
where we put in ${\rm RHS}(\tau)$ all the terms that involve
interactions (i.e., non-linearities in $X$). The ${\rm RHS}$ depends
on the past values of the field and of the propagators, but we shall
not write them for compactness. Also, in the case of the propagators,
$X$ depends on a second time argument $\tau'$, which is not written
explicitly here.

Writing ${\bs X}\equiv \big({{X}\atop{\tau\partial_\tau X}}\big)$, it
is easy to check that the solution of this differential equation over
one timestep $\tau_{n}\to \tau_{n+1}$ reads
\begin{align}
  {\bs X}(\tau_{n+1})
  =
  {\bs L}(\tau_{n+1},\tau_{n}){\bs X}(\tau_{n})
  +
  \int_{\tau_{n}}^{\tau_{n+1}} d\tau''\, \tau''\; {\bs L}(\tau_{n+1},\tau'')\left({{0}\atop{{\rm RHS}(\tau'')}}\right),
  \label{eq:genSol}
\end{align}
where we have defined
\begin{align}
  {\bs L}(\tau,\tau')=
  \frac{i\pi }{4}
  \begin{pmatrix}
    m_\perp \tau'\big[H_{i\nu}^{(1)}(m_\perp \tau)\dot{H}_{i\nu}^{(2)}(m_\perp \tau')
    &
    H_{i\nu}^{(2)}(m_\perp \tau)H_{i\nu}^{(1)}(m_\perp \tau')
    \\
    -H_{i\nu}^{(2)}(m_\perp \tau)\dot{H}_{i\nu}^{(1)}(m_\perp \tau')\big]
    &
    -H_{i\nu}^{(1)}(m_\perp \tau)H_{i\nu}^{(2)}(m_\perp \tau')
    \\
    &\\
    m_\perp^2\tau\tau'\big[\dot{H}_{i\nu}^{(1)}(m_\perp \tau)\dot{H}_{i\nu}^{(2)}(m_\perp \tau')
      &
      m_\perp {\tau}\big[\dot{H}_{i\nu}^{(2)}(m_\perp \tau)H_{i\nu}^{(1)}(m_\perp \tau')
      \\
      -\dot{H}_{i\nu}^{(2)}(m_\perp \tau)\dot{H}_{i\nu}^{(1)}(m_\perp \tau')\big]
    &
    -\dot{H}_{i\nu}^{(1)}(m_\perp \tau)H_{i\nu}^{(2)}(m_\perp \tau')\big]
      \\
    \end{pmatrix}.
\end{align}
Given precomputed tables of values for the Hankel functions, the matrix ${\bs L}$
would be known with machine accuracy. 

In a practical algorithm, the function ${\bs X}(\tau)$ (and therefore
${\rm RHS}(\tau)$) is known only at discrete values  of the proper
time. In the determination of ${\bs X}(\tau_{n+1})$ using
eq.~(\ref{eq:genSol}), we therefore do not know ${\rm RHS}(\tau'')$ at
any point $\tau''$ in the range $[\tau_n,\tau_{n+1}]$, except for the
lower boundary $\tau''=\tau_n$. This problem may be circumvented by
using an extrapolative quadrature rule, that uses a few of the points
preceding the integration range. We list here three such quadrature
formulas, in order of increasing accuracy:
\begin{align}
  \int_{\tau_n}^{\tau_{n+1}}d\tau'' \,A(\tau'')
  &=
  \varepsilon\,A(\tau_n)+{\cal O}(\varepsilon^2 \big|A'\big|),\nonumber\\
  &=
  \varepsilon\,\Big[\frac{3}{2}A(\tau_n)-\frac{1}{2}A(\tau_{n-1})\Big]+{\cal O}(\varepsilon^3 \big|A''\big|),\nonumber\\
  &=
  \varepsilon\,\Big[\frac{23}{12}A(\tau_n)-\frac{16}{12}A(\tau_{n-1})+\frac{5}{12}A(\tau_{n-2})\Big]+{\cal O}(\varepsilon^4\big|A'''\big|),
\end{align}
where $\varepsilon\equiv \tau_{n+1}-\tau_n$ is the timestep (assumed
to be constant). These formulas are exact when $A(\tau)$ is a
polynomial of degree $0$, $1$, or $2$, respectively. Note that, in
order to use the two-step (resp. three-step) method, we must store the
values of $A$ (i.e., of ${\rm RHS}$) obtained in the previous step
(resp. the previous two steps) of the time evolution.

\paragraph{Diagonal propagators:}
\begin{figure}[htbp]
  \centering
  \includegraphics[width=80mm]{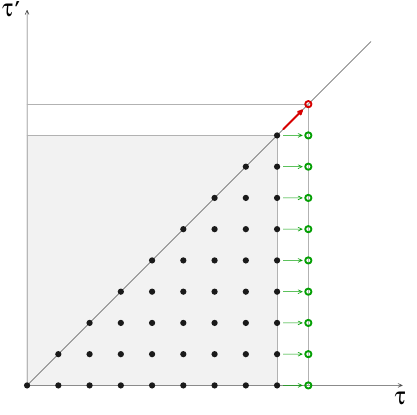}
  \caption{\label{fig:evol}Illustration of the evolution of the system by one timestep.}
\end{figure}
The evolution algorithm we have just described allows to evolve the
propagators by one timestep in the $\tau$ direction, at fixed
$\tau'$, as illustrated by the green arrows in Figure
\ref{fig:evol}. Before this timestep, all the propagators in the area
shaded in gray have already been computed (because $F$ is symmetric
under the exchange of the two times, and $\rho$ antisymmetric, we only
store the propagators corresponding to pairs $(\tau,\tau')$ below
the diagonal in this picture). The method outlined above in this
section therefore gives all the propagators (and their derivative with respect to
$\tau$) at the points shown in green.

However, this method cannot give the propagators at
$(\tau+\varepsilon,\tau+\varepsilon)$, as this requires doing a
timestep in both time variables simultaneously. Note first that obtaining
$\rho$ and its derivative at equal times is trivial, since we have
\begin{align}
  \rho(\tau,\tau)=0,\quad
  \tau \partial_{\tau}\rho(\tau,\tau')\empile{=}\over{\tau=\tau'}-1,
\end{align}
according to the canonical commutation relations. Therefore, we only need to
determine $F$ on the diagonal. This can be achieved by generalizing
the previous algorithm as follows. Firstly, let us write the equations of motion for $F$ in the $\tau$ and $\tau'$ directions as
\begin{align}
  & \Big(\partial_{\tau}^2+\tau^{-1}\partial_{\tau}\Big)F+\omega^2 F = {\rm RHS},\nonumber\\
  & \Big(\partial_{\tau'}^2+{\tau'}^{-1}\partial_{\tau'}\Big)F+{\omega'}^2 F = {\rm RHS}',
  \label{eq:eom3}
\end{align}
with
\begin{align}
  \omega^2\equiv m^2+p_\perp^2+\frac{\nu^2}{\tau^2},
\end{align}
(and a similar definition for ${\omega'}^2$)
and where ${\rm RHS}$ collects all interaction terms. Then, introduce the
following $2\times 2$ matrix:
\begin{align}
  {\bs F}\equiv
  \begin{pmatrix}
    F & \tau'\partial_{\tau'} F\\
    \tau\partial_\tau F & \tau\tau'\partial_\tau \partial_{\tau'} F\\
  \end{pmatrix}.
\end{align}
The evolution along the diagonal is solved by
\begin{align}
  &{\bs F}(\tau_{n+1},\tau_{n+1})
  =
  {\bs L}(\tau_{n+1},\tau_n){\bs F}(\tau_n,\tau_n){\bs L}^t(\tau_{n+1},\tau_n)
  \nonumber\\
  &
  \;+\int\limits_{\tau_n}^{\tau_{n+1}} \!\!d\tau''\,\tau''\,
  {\bs L}(\tau_{n+1},\tau'')
  \begin{pmatrix}
    0 & {\rm RHS}(\tau'',\tau'')\\
    {\rm RHS}(\tau'',\tau'') & 2\tau''\partial_{\tau''} {\rm RHS}(\tau'',\tau'')\\
  \end{pmatrix}
  {\bs L}^t(\tau_{n+1},\tau''),
\end{align}
which allows to generalize the method introduced earlier to simultaneous moves
in $\tau$ and $\tau'$ (note that the terms ${\rm RHS}$ and ${\rm RHS}'$
in the two equations (\ref{eq:eom3}) are equal when $\tau=\tau'$). Note that this evolution along the diagonal requires that we keep track of the double derivative $\partial_\tau\partial_{\tau'}F$ at $\tau=\tau'$.

\section{Results}
\label{sec:Results}

\subsection{Setup of the computation}
\label{sec:setup_calc}

Simulations with the 2PI effective action are initialized by specifying the statistical function \(F\) and the spectral function \(\rho\), as well as their derivatives, at an initial time \(\tau_0\). In this work we choose a simple initialization where \(\rho\) is the free spectral function, \(\rho(\tau_0, \tau_0; p_{\perp}, \nu) = 0\) and \(\partial_{\tau} \rho(\tau, \tau'; p_{\perp}, \nu)\big|_{\tau=\tau'=\tau_0} = -\tfrac{1}{\tau_0}\), as follows from the commutation relations of a scalar field.\footnote{For further details see App. \ref{App:quasiparticle}. The factor  \(1/\tau_0\) comes from the determinant of the metric in our coordinate system.} Furthermore, \(F\) is the free statistical function with an initial occupation density  
\(f(p_{\perp},\nu)\), and takes the form 
\beq
\label{Eq:initialF}
F(\tau_0, \tau_0; p_{\perp},\nu) = \left(\frac{1}{2} + f(p_{\perp},\nu)\right) \frac{\pi}{2} e^{-\pi \nu} \left| H_{i\nu}^{(1)}(m_\perp \tau_0)\right|^2,
\eeq
in our coordinate system. 
In this expression \(m_\perp = \sqrt{p_{\perp}^2 + m^2}\) is the transverse mass and the factor \(1/2\) is the vacuum contribution to the occupation density. Similar expressions for the first and second time derivatives, \(\partial_{\tau} F(\tau, \tau^{\prime}; p_{\perp}, \nu) \big\rvert_{\tau=\tau^{\prime} = \tau_0 } \) and \(\partial_{\tau} \partial_{\tau^{\prime}} F(\tau, \tau^{\prime}; p_{\perp}, \nu) \big\rvert_{\tau=\tau^{\prime} = \tau_0 }\), are needed for the initialization. These can be found in App. \ref{app:bareprop}, as well as a derivation of Eq. \eqref{Eq:initialF}. We set the 1-point function \(\varphi = 0\) initially which guarantees that \(\varphi(\tau) = 0\) at all times.

For the initial occupation density, we choose a Gaussian distribution
\beq
\label{Eq:f0}
f(p_{\perp}, \nu) = f_0 \; e^{-p_{\perp}^2/Q^2} e^{-\nu^2/\beta^2}.
\eeq
The prefactor is set to $f_0=12.5$, the 
momentum scale to \(Q = 4.0\) (in lattice units) and \(\beta = Q\tau_*\). Since $p_z\approx \nu/\tau$, the time $\tau_*$ (set to $\tau_*=1$ in lattice units) is the time at which this distribution would be isotropic  in the absence of scatterings (but, without scatterings, the system would become anisotropic at later times due to the longitudinal expansion).
This initialization sets the typical transverse momentum to \(Q\). Therefore, $Q$ plays the same role as the saturation scale $Q_s$ in the color-class-condensate framework, and some intuition for the scales of our problem can be obtained by identifying $Q$ and \(Q_s\).

Our initialization is simple and gives a benchmark for 2PI simulations in a longitudinally expanding medium. However, it makes a number of assumptions. Most importantly, the initial propagators describe excitations with only a vacuum mass \(m\). Once interactions are turned on and the system starts evolving, the excitations immediately acquire an additional thermal mass, as well as corrections to the vacuum mass due to vacuum fluctuations. This produces a mass quench, i.e. a sudden change in the effective mass, which can lead to spurious oscillations in the time evolution.  A general solution to this issue is to include $n$-point sources at the initial time, which effectively make the initial state closer to the subsequent evolving system and thus reduce this spurious mass quench \cite{Garny:2009ni}.
Such $n$-point sources furthermore allow for more complicated correlations in the initial conditions. 
 In our current work, the coupling is small enough that the initial mass quench does not pose serious problems, as the resulting oscillations are quickly damped by scattering and the simulations are stable.
 
We initialize the system at a very early time 
\(\tau_0 Q=0.016\).
The precise value of \(\tau_0\) is unimportant as long as \(\tau_0 Q\ll 1\) because the system is dominated by free streaming at these very early times. We furthermore set the mass to 
\(m= 0.625 Q\).
This relatively large mass allows us to do calculations at slightly stronger coupling. The lattice has sizes \(N_{\eta} = 1536\) and \(N_{\perp} = 512\).  

We choose the maximum value of \(p_{\perp}\) to be 
\(P = 6.7 Q\)
and we choose \(a_{\eta}\) so that the maximum value of \(\nu\) is \(\nu_{\mathrm{max}} = 300\). Such a large value for \(\nu_{\mathrm{max}}\) is needed since the maximal $p_z$ representable on our lattice decreases with time as $p_{z,\rm{max}}\approx \nu_{\rm max}/\tau$. We store the \(N_{\tau} = 300\) most recent time steps for evaluating the memory integrals, with a step size 
\(\Delta \tau \,Q = 0.048\).

Our goal is to explore the onset of isotropization for different values of the coupling strength \(g^4\). The 2PI effective action, when truncated in the number of loops, puts an upper limit on \(g^4\) because the thermal mass becomes negative for too large values of the coupling, see e.g. \cite{Arrizabalaga:2005tf}. This is most easily understood by recalling that the thermal mass  in massless \(\phi^4\) perturbation theory in thermal equilibrium in a fixed box is \cite{deGodoyCaldas:2001mb}
\beq
m_{\mathrm{th; \;eq}}^2 = \frac{g^2 T^2}{24} \left[ 1 - 3 \left(\frac{g^2}{24\pi^2} \right)^{1/2}\right]. 
\eeq
This expression is obtained by including tadpole contributions and solving the gap equation self-consistently.
This expression implies that the total thermal mass becomes negative at \(g^4 \gtrsim 64\pi^4/9 \approx 690\), signalling that the expansion in the number of loops has broken down. A similar behaviour is seen in our full 2PI calculation despite two-loop contributions and an expanding background: for too large coupling, one obtains a negative thermal mass and the calculation is unstable, see also Sec. \ref{sec:th_mass}. For this reason, we restrict our simulations to \(g^4 \leq 500\). 
This breakdown of the theory at larger coupling is not visible in kinetic theory simulations since they do not include a full 2-loop thermal mass. This is another example of the limitations of kinetic theory which contains  less information than a 2PI calculation and can thus be used erroneously at too large coupling.

Our maximal coupling of \(g^4 = 500\) corresponds to a system which is not yet strongly interacting. This can be seen by noting that the ratio of shear viscosity to entropy density at this coupling in \(\phi^4\) theory is \(\eta /s  = 3033.5/g^4 \approx 6\) \cite{Jeon:1994if,Jeon:1995zm}.
This ratio is still much larger than the strong coupling limit \(\eta/s = 1/4\pi \approx 0.08\) of SUSY $N=4$ Yang-Mills theory.

\subsection{Isotropization}

Our goal is to study isotropization in an  expanding medium described by quantum field theory (as approximated by truncating the 2PI effective action at order $g^4$).
A typical observable to study isotropization is the stress-energy tensor \(T^{\mu\nu}\) or more precisely the ratio of the longitudinal and transverse pressures \(P_L / P_T\). The stress-energy tensor coming from the 2PI effective action has a complicated dependence on UV cutoffs \(\Lambda\). The unrenormalized version depends on the cutoff as \(\Lambda^4\). A simple subtraction of the vacuum stress-energy tensor \(T_0^{\mu\nu}\) still leaves a \(\Lambda^2\) dependence which comes from overlapping medium and vacuum contributions. Its subtraction requires a detailed analysis that we postpone to future work. 

In this work, we focus on quantities that have less sensitivity to the UV regularization. One such quantity is the occupation density \(f(p_{\perp},\nu; \tau)\).  To define the occupation density, we assume a quasiparticle picture where the propagators approximately take the form
\begin{align}
\label{Eq:F_quasiparticle}
  {F}(\tau,\tau';p_\perp,\nu)
  &=
  \frac{\pi (\tfrac{1}{2}+f(p_\perp,\nu;\tau))}{4}
  \!\Big(\!
  H_{i\nu}^{(1)}(\mbar_\perp \tau)H_{i\nu}^{(2)}(\mbar_\perp \tau')
  \nonumber\\
  &\qquad\qquad\qquad\qquad+\!
  H_{i\nu}^{(2)}(\mbar_\perp \tau)H_{i\nu}^{(1)}(\mbar_\perp \tau')
  \!\Big),
\end{align}
\begin{align}
\label{Eq:rho_quasiparticle}
  {\rho}(\tau,\tau';p_\perp,\nu)
  &=
  \frac{i\pi}{4}
  \!\Big(\!
  H_{i\nu}^{(1)}(\mbar_\perp \tau)H_{i\nu}^{(2)}(\mbar_\perp \tau')
  \!-\!
  H_{i\nu}^{(2)}(\mbar_\perp \tau)H_{i\nu}^{(1)}(\mbar_\perp \tau')
  \!\Big).
\end{align}
where $\mbar_{\perp} = \sqrt{p_\perp^2 + m^2 + M^2(p_\perp,\nu;\tau)}$ with $m^2$ the bare vacuum mass and $M^2(p_{\perp},\nu; \tau)$ the corrections to the mass. Note that \(M^2\) may be momentum dependent, because it includes not only the tadpole contribution but also self-energies beyond one-loop.\footnote{This term can furthermore include vacuum corrections to the bare mass.} In the quasiparticle picture, the propagators have the same form as in free streaming except that the occupation density and the medium corrections to the mass vary with time. 
Furthermore, these variations are assumed to be slower than the oscillations of the Hankel functions, whose typical time scale  is controlled by $1/\mbar_{\perp}$.

We emphasize that our 2PI simulations solve the equations of motion in full generality and  do not assume a quasiparticle picture. Rather, we fit the full solutions for \(F\) and \(\rho\) to the quasiparticle ansatz, Eqs. \eqref{Eq:F_quasiparticle} and \eqref{Eq:rho_quasiparticle}, in order to extract an occupation density and a thermal mass. The details of this procedure are described in App.  \ref{App:quasiparticle}. 
The extracted occupation density will not be the same as for a kinetic theory calculation since the evolution equations are different.

\begin{figure}[htbp]
  \centering
  \includegraphics[width=1.0\textwidth]{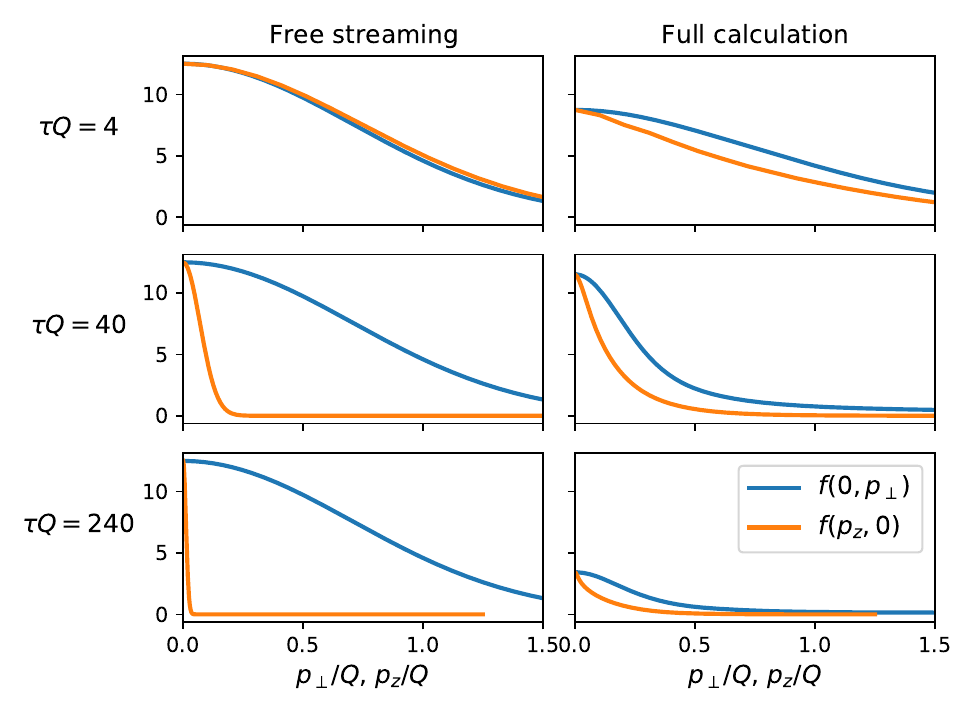}
  \caption{ Occupation density \(f(p_{\perp},p_{z})\) extracted from 2PI calculations using the quasiparticle ansatz in Eq. \ref{Eq:F_quasiparticle}.  We plot the  occupation density along the lines \(p_{\perp}=0\) and \(p_z = 0\) for three different times and for both free streaming (i.e., $g^4=0$) and a full calculation. The full calculation shows signs of isotropization as the orders of magnitude of \(p_{\perp}\) and \(p_z\) stay the same while in the free streaming case \(p_z\) drops very fast due to the expansion. The full calculation has \(g^4 = 500\), \(m/Q=0.625\).}
  \label{fig:momdistr}
\end{figure}

In Fig. \ref{fig:momdistr} we  plot the occupation density \(f(p_{\perp},p_z)\) at fixed times.   We compare a full 2PI calculation at \(g^4 = 500\) with free streaming at \(g^4 = 0\).  In free streaming, particles are at constant \(\nu\) due to expansion and thus the longitudinal momentum \(p_z = \nu/ \tau\) decreases like \(1/\tau\). On the contrary, in the full calculation the longitudinal momentum decreases much more slowly and the typical values of \(p_{\perp}\) and \(p_z\) remain comparable in magnitude at all times. This gives signs of isotropization in the 2PI framework. We furthermore see that in the full 2PI calculation the overall magnitude of \(f(p_{\perp},p_z)\) decreases with time. This is simply because the number density per unit volume \(n = \int \frac{d^2 p_{\perp}}{(2\pi)^2} \frac{d p_z}{2\pi} f(p_{\perp},p_z)\) decreases with time (like $\tau^{-1}$ if number changing processes are negligible) due to the longitudinal expansion, while the support in $(p_z,p_\perp)$ of the particle distribution does not shrink much.

\begin{figure}[htbp]
  \centering
  \includegraphics[width=1.0\textwidth]{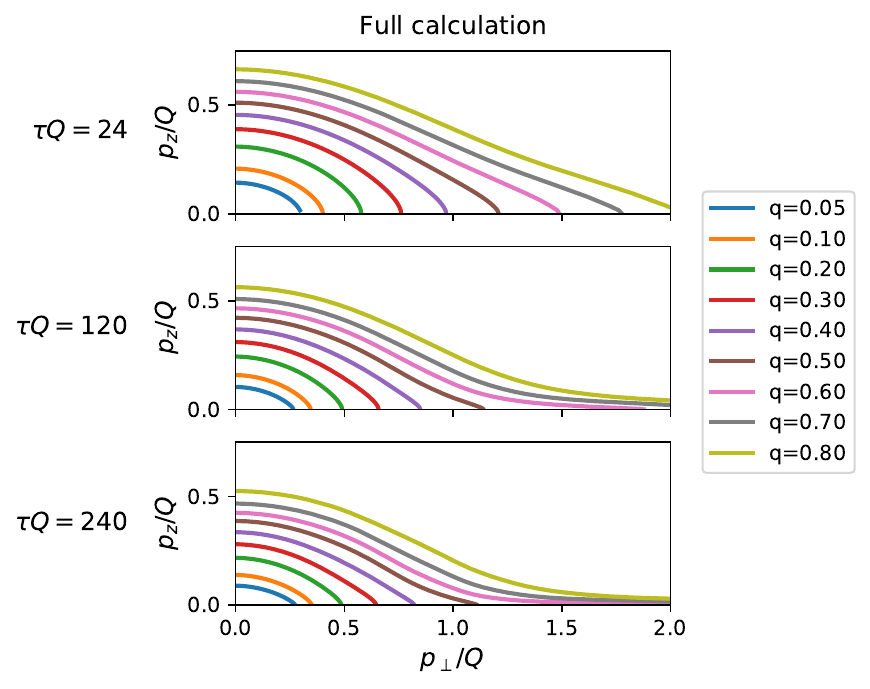}
  \caption{Contours in the \(\left(p_{\perp}, p_z \right)\) plane along which the occupation density \(f(p_{\perp},p_{z})\) takes a constant value. The different contours are chosen so that a fraction \(q\) of the total number of particles are inside the contour in momentum space. Shown is a full 2PI calculation with  \(g^4 = 500\), \(m/Q=0.625\).}
  \label{fig:contourpanel}
\end{figure}

To study isotropization more carefully, we consider contours in the phase space \((p_{\perp},p_z)\) along which the occupation density \(f(p_{\perp},p_z)\) takes fixed values \(f_*\). 
Such contours are circular for  an isotropic occupation density  and become more elongated as the anisotropy increases. 
We choose the values \(f_*\) such that a certain fraction \(q\) of the particles are inside the contour, i.e.
\beq
\int_{\rm inside\;\; contour} \frac{d^2p_{\perp}}{(2\pi)^2} \;\frac{d\nu}{2\pi} \; f(p_{\perp},\nu)= q \,\bar{n}
\eeq
where \(0\leq q \leq 1\) and \(\bar{n} = \int \frac{d^2p_{\perp}}{(2\pi)^2} \frac{d\nu}{2\pi} f(p_{\perp},\nu)\) is the total number of particles per unit rapdity and transverse area.  The results for contours at constant value of the occupation number \(f(p_{\perp},\nu)\) can be seen in Fig. \ref{fig:contourpanel}. We show the contours at different times and for different values of \(q\).
For the times we look at, there is a slight reduction in typical values of \(p_z\) while \(p_{\perp}\) stays roughly constant with the tail of the distribution going to higher \(p_{\perp}\). 
The typical values of \(p_z\) are lower than typical values of \(p_{\perp}\) but their ratio stays more or less constant showing that the system is slowly isotropizing despite the rapid expansion.

\begin{figure}[htbp]
  \centering
  \includegraphics[width=1.0\textwidth]{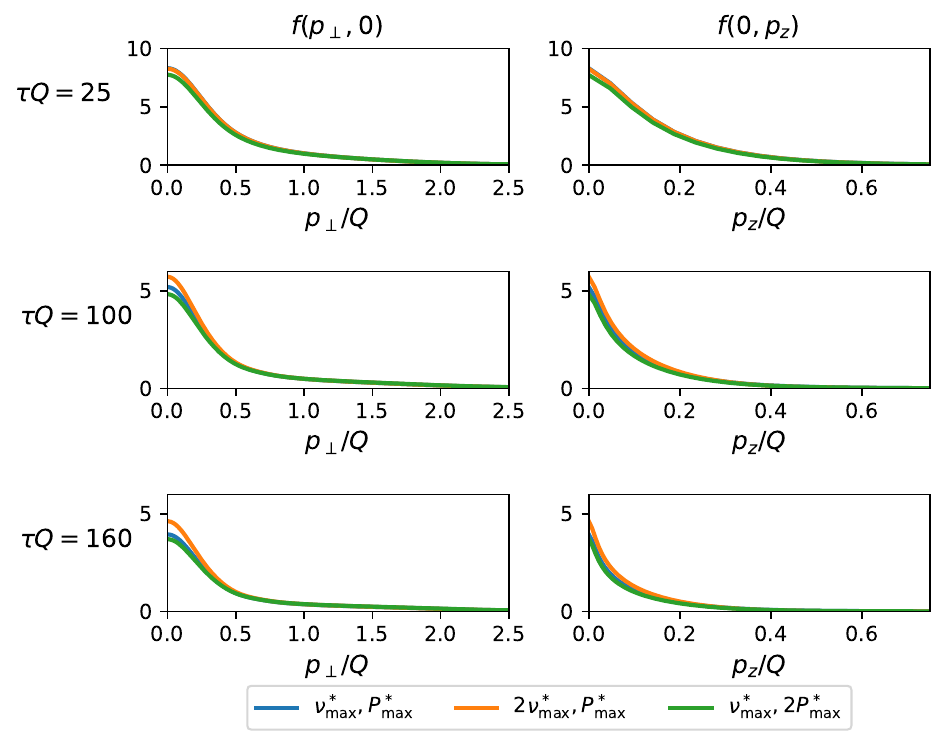}
\caption{ The momentum distribution extracted in calculations with different grid sizes. The first grid size shown is the one used in the rest of the paper, \(\nu_{\max} = \nu_{\mathrm{max}}^*, P_{\mathrm{max}} = P_{\mathrm{max}}^*\) where \(\nu_{\mathrm{max}}^* = 300\) and \(P_{\mathrm{max}}^*  = 6.8 Q\). The second grid  has twice the extent  in the longitudinal direction, \(\nu_{\max} = 2 \nu_{\max}^*, P_{\max} = P_{\max}^*\), and the third one has twice the extent in the transverse direction \(\nu_{\max} = \nu_{\max}^*, P_{\max} =2  P_{\max}^*\). Changing the extent of the grid mostly matters at low \(p_{\perp}\) and \(\nu\) and gives at most a \(20\,\%\) difference.}
\label{fig:mom_distr_momcutoff}
\end{figure}

Finally, as a check of our calculation we vary the size of the lattice and verify how much the occupation density \(f(p_{\perp},\nu)\) is modified. This checks the robustness of our 
minimal renormalization where only power-law divergences in the tadpole are subtracted, as explained in Sec. \ref{sec:renorm}. In Figs. \ref{fig:mom_distr_momcutoff} we show \(f(p_{\perp},\nu=0)\) and \(f(p_{\perp}=0,\nu)\), for our usual grid where the maximum value of \(p_{\perp}\) is set to \(P=6.8 Q\) and the maximum value of \(\nu\) is set to \(\nu_{\rm max}= 300\), as well as for grids with twice the value of \(P\) or twice the value of \(\nu_{\rm max}\).  The dependence is fairly mild, with the most pronounced difference being at low \(p_{\perp}\) and \(\nu\). The maximal difference is around \(20\,\%\) at lower momenta. The UV tail also differs between the calculations making it difficult to calculate quantities such as \(\langle p_{\perp} \rangle =  \int_{p_{\perp},\nu} p_{\perp} f(p_{\perp},\nu) \) and \(\langle p_{z} \rangle =  \int_{p_{\perp},\nu} \left|p_{z}\right| f(p_{\perp},\nu) \) which  depend more heavily on the UV behaviour. This makes it difficult to describe the rate of isotropization more quantitatively in our current setup.
We will return to this issue in future work after having implemented more rigorous renormalization which gives cutoff a more accurate independence in the UV.

\begin{figure}[htbp]
  \centering
  \includegraphics[width=0.7\textwidth]{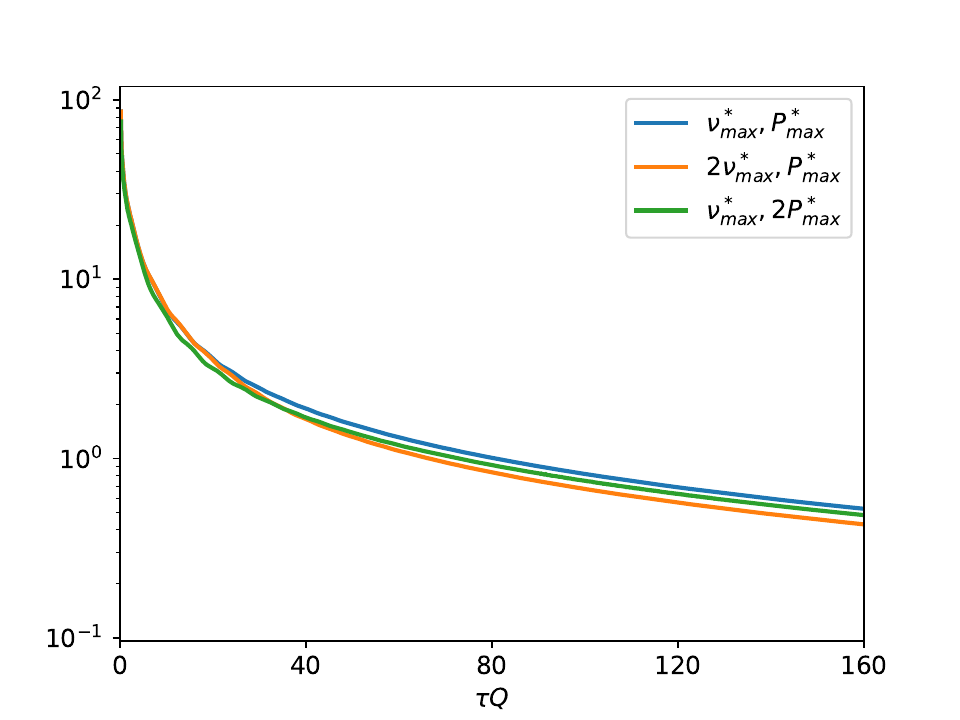}
  \caption{The renormalized tadpole defined in Eq. \eqref{Eq:ren_tadpole} for three different grid sizes (the same as in Fig. \ref{fig:mom_distr_momcutoff}). By doubling the extent of the grid in either the longitudinal or transverse momentum directions, the renormalized tadpole varies at most by \(20\,\%\) at \(\tau Q=160\).}
  \label{fig:tadpole_momcutoff}
\end{figure}

As an additional check on our renormalization procedure,  we show in Fig. \ref{fig:tadpole_momcutoff} the renormalized tadpole
\beq
\label{Eq:ren_tadpole}
\tfrac{g^2}{2}\int \frac{dp_\perp p_\perp\,d\nu}{(2\pi)^2}\, \left[ F(\tau,\tau,p_\perp,\nu) - F_0(\tau,\tau,p_\perp,\nu) \right]
\eeq
for three grids in momentum space with different cutoffs.
The difference is around \(20\,\%\) at \(\tau Q = 160 \). This residual cutoff dependence is because in Eq. \eqref{Eq:ren_tadpole} we have only removed order $g^2$ divergences, but not higher-order divergences.

\subsection{Thermal mass}
\label{sec:th_mass}

The quasiparticle ansatz in Eq. \eqref{Eq:F_quasiparticle} not only gives the occupation density of particles but also their dispersion relation. The dispersion relation is determined by the mass correction \(M^2(p_{\perp},\nu; \tau)\) which changes with time and also depends on the momentum of quasiparticles. This thermal mass can  be anisotropic, meaning that it depends not only on the magnitude of the momentum \(p= \sqrt{p_{\perp}^2 + p_z^2}\) but also on its orientation (or, equivalently, depends separately on $p_\perp$ and $\nu$). All of this gives complicated dispersion relations in our 2PI calculations. Our method for extracting the thermal mass is described in App. \ref{App:quasiparticle}.

\begin{figure}[htbp]
  \centering
  \includegraphics[width = 0.7\textwidth]{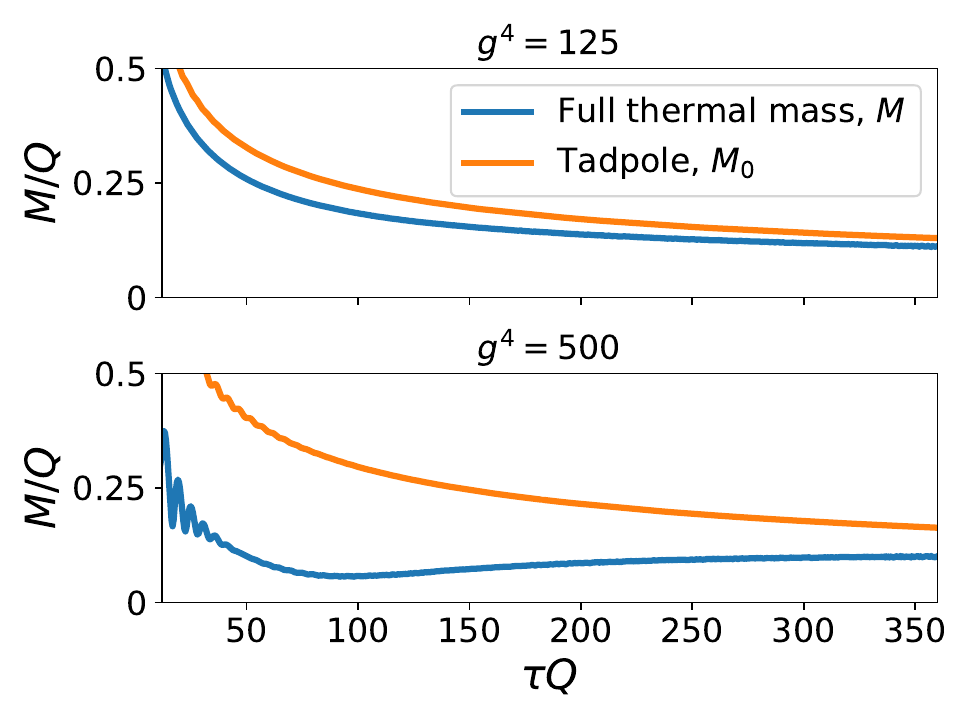}
  \caption{The thermal mass \(M(p_{\perp},p_z)\) at momentum \(p_{\perp} = 0\), \(p_z = 0\) as a function of time at two different couplings. Shown is both the full thermal mass and the tadpole contribution at \(g^2\), defined in Eq. \ref{Eq:M2_unrenorm1}. Interestingly, the full thermal mass is lower at higher coupling due to the (negative) contribution of order \(g^4\) becoming more important.}
  \label{fig:th_mass_time}
\end{figure}

In Fig. \ref{fig:th_mass_time} we show the thermal mass \(M(p_{\perp},\nu;\tau)\) for \(p_{\perp}=0\) and \(\nu=0\) as a function of time \(\tau\) for two values of the coupling, \(g^4 = 125\) and \(g^4 = 500\). As expected, the thermal mass decreases with time as the density of particles goes down. Interestingly, the full thermal mass is lower at higher coupling. This is explained by the relative importance of the tadpole contribution at order \(g^2\) which is positive and the contribution at order \(g^4\) which is contained in the memory integrals and which is negative. At lower coupling the tadpole contribution dominates but at higher coupling the \(g^4\) contribution becomes more important and lowers the total thermal mass.  This also shows that in our truncation scheme the full thermal mass may become negative if the coupling is too high, signalling that the approximation of truncating \(\Phi\) at three loops has broken down.
This effect is normally not included in kinetic theory calculations despite the fact that they use the same truncation scheme. One needs to be careful not to use kinetic theory outside this range of applicability.

We now turn to the momentum dependence of the dispersion relation which is a crucial ingredient to understand isotropization in a longitudinally expanding medium. This can  be seen, e.g., in the  ``bottom-up'' scenario in QCD where, in the final stage, high-energy particles are quenched in a medium formed by soft particles \cite{Baier:2000sb}. (More precisely, the high-energy particles receive kicks from the soft particles which stimulate repeated medium-induced radiation.) The details of this quenching depend on the dispersion relations of the soft particles. Normally, the soft particles are assumed to have already been isotropized due to repeated two-to-two scattering. However, this isotropization of the soft sector has not been shown unambigously in numerical calculations as kinetic theory calculations do not treat the soft sector dynamically but rather assume it to be isotropic and classical-statistical simulations work at extremely small coupling and cannot be trusted up until the final stage of the ``bottom-up'' scenario. A dynamical understanding of the soft sector at realistic coupling is made even more important by the potential presence of instabilities which can alter the mechanism of isotropization. 

\begin{figure}[htbp]
  \centering
  \includegraphics[width = 1.0\textwidth]{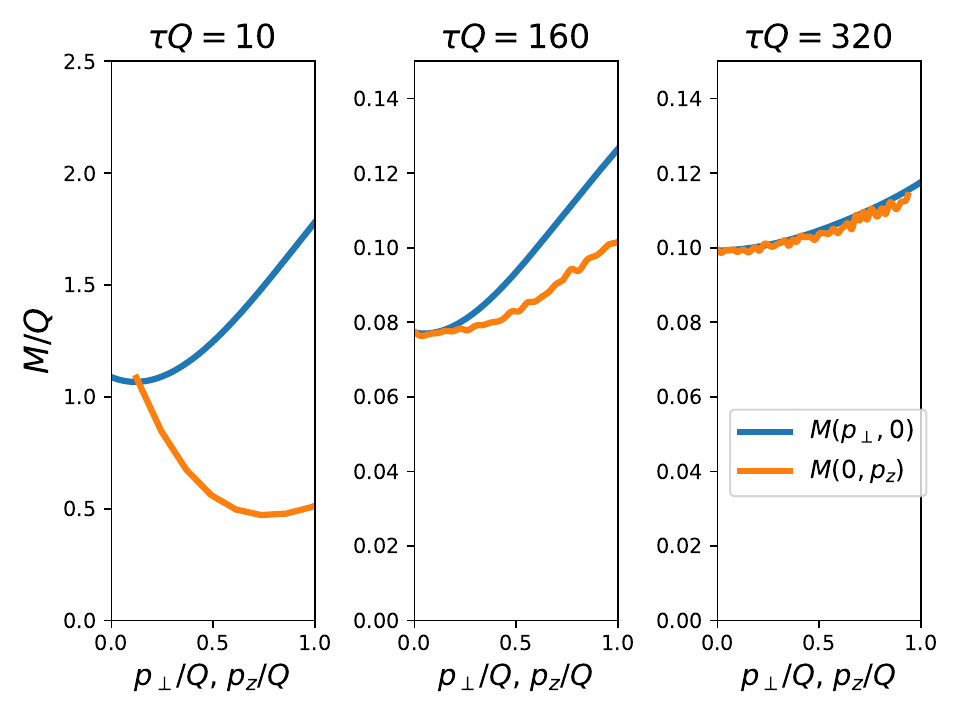}
  \caption{ The momentum dependence of the thermal mass at three different times. We show \(M(p_{\perp},0)\) where \(p_z= 0\) is kept fixed and \(M(0,p_z)\) where \(p_{\perp}= 0\) is kept fixed. At full isotropy these two curves become identical. The figure shows that initially the two curves are far from each other so that the dispersion relation is heavily anisotropic but that nearly full isotropy is reached around \(\tau Q=320 \).}
  \label{fig:th_mass}
\end{figure}

It is beyond the scope of this work to verify whether the soft sector in heavy-ion collisions isotropizes before it breaks up energetic particles. However, our work in scalar field theory can shed some light on this question. In Fig. \ref{fig:th_mass}, we show the dependence of the thermal mass \(M\) on momentum at three different times \(\tau Q=10\), \(\tau Q=160\) and \(\tau Q=320\) in a calculation with \(g^4 = 500\). Specifically, we plot \(M(p_{\perp},0)\) and \(M(0,p_z)\). These two curves become identical at full isotropy. Fig. \ref{fig:th_mass} shows that initially the thermal mass is strongly anisotropic but as time goes on it isotropizes, reaching nearly full isotropy at around \(\tau Q=320\). This effect is not because the momentum-independent tadpole contribution dominates since this contribution is only responsible for half the mass at this time, see Fig. \ref{fig:th_mass_time}.
Rather it 
truly reflects isotropization of the soft sector.
 This isotropization of the thermal mass is a faster process than the isotropization of the occupation density of hard particles. However, it is far from  instantaneous suggesting that a full description of a scalar medium  
 needs to take into account the non-equilibrium anisotropic contribution to the disperion relation.

\subsection{Number-changing processes}
Another major difference between the 2PI effective action and kinetic theory is that the 2PI effective action allows for number-changing processes, even when the 2PI functional $\Phi$ is truncated at three loops. This is 
due to the possible off-shellness of the quasi-particle excitations (i.e., to the fact that the exact spectral function is not proportional to a delta function, but has a non-zero width instead), which allows for \(1\rightarrow 3\) and  \(3\rightarrow 1\) processes in addition to the usual \(2 \rightarrow 2\) processes (whose kinematics is the only one that is possible for on-shell particles in kinetic theory). Number-changing processes were analyzed in detail in a static system in \cite{Juchem:2003bi} and shown to be responsible for chemical equilibration.

It is important to quantify how much the number-changing processes contribute in 2PI calculations in an expanding system. A straightforward way to estimate the change in number density is to assume that \(F\) is described by the quasiparticle ansatz in Eq. \eqref{Eq:F_quasiparticle} with occupation density \(f(p_{\perp},\nu)\) and then to calculate as a function of time the corresponding number density 
\beq
\overline{n}(\tau) = \int \frac{d^2 p_{\perp}}{(2\pi)^2} \int \frac{d\nu}{2\pi}\; f(p_{\perp},\nu; \tau). 
\eeq
Here, \(\overline{n}\) is the total number of particles in a unit area in the transverse plane and in one unit in rapidity. This quantity is constant in kinetic theory, \(d\overline{n} / d \tau = 0\).\footnote{A more common form of number conservation in a longitudinally expanding medium reads \(\frac{d (\tau n)}{d\tau} = 0\) where \(n\) is the number of particles per unit volume $d^3\x$. The two are simply related through \(n = \overline{n} / \tau\) giving that \(\overline{n}\) is constant in time in kinetic theory.
}

\begin{figure}[htbp]
  \centering
  \includegraphics[width=0.9\textwidth]{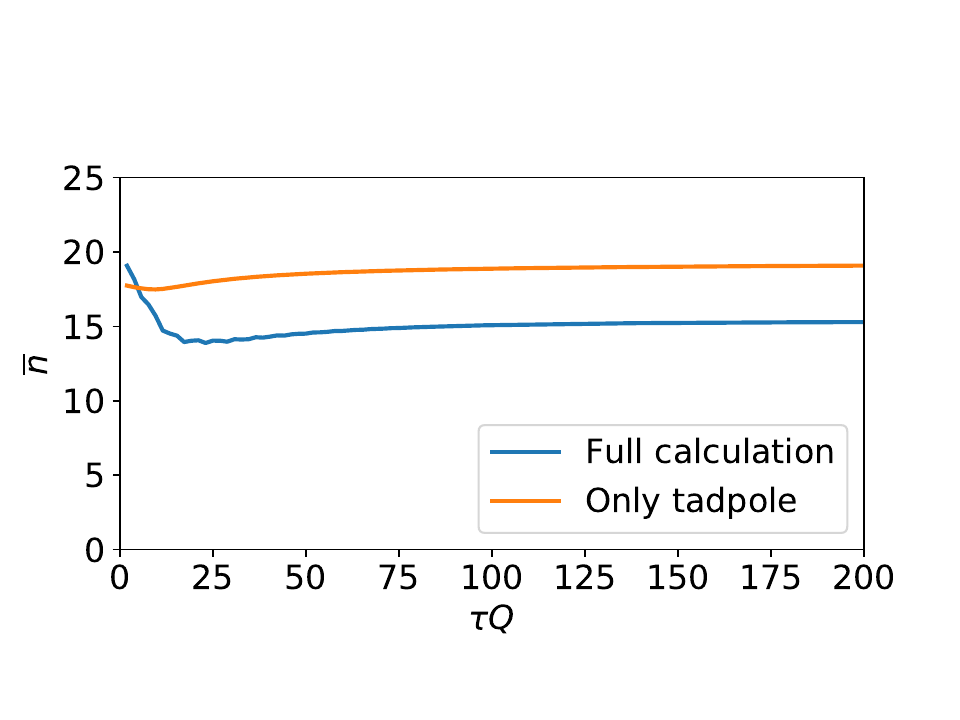}
  \caption{Number density \(\overline{n}\) in a unit rapidity and transverse area as a function of time. Both a full 2PI calculation and a calculation that only includes mean-field effects given by the tadpole are shown. In the full calculation there is a roughly \(20\,\%\) reduction in number density due to number-changing processes.} 
  \label{fig:number}
\end{figure}

In Fig. \ref{fig:number}, we show the evolution of \(\overline{n}(\tau)\) with time in our 2PI calculations at \(g^4 = 500\). Shown are both a full calculation and 
one that only includes the tadpole and not the scatterings coming from the sunset diagram. As expected, \(\overline{n}\) stays nearly constant in the absence of scatterings. Deviations from a constant behavior in the full calculation may be attributed to the fact that this 2PI calculation deviates from the simple quasiparticle ansatz. In contrast, in the full calculation \(\overline{n}\) decreases substantially at early times so that the final number density is \(20\,\%\) below the initial number density. This shows that number-changing processes are important for quantitative predictions in scalar field theory, especially at early times and high occupation density.

\begin{figure}[htbp]
  \centering
  \includegraphics[width=100mm]{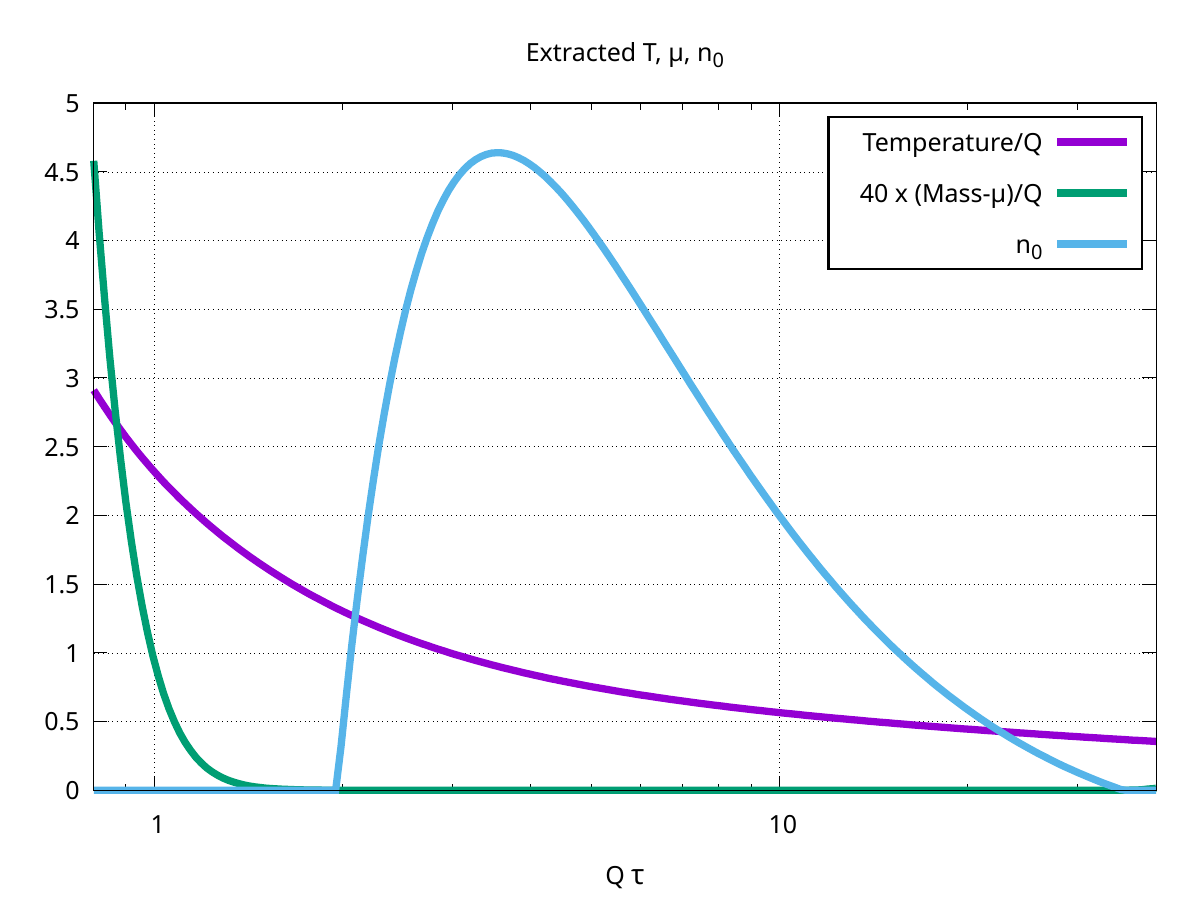}
  \caption{Temperature, chemical potential and the density of the condensate that one would get by instant thermalization, with the number and energy kept fixed. The extracted values show that the medium we simulate is overoccupied.}
  \label{fig:Tmu}
\end{figure}

One can easily understand why the number density in the full 2PI calculation decreases at early times. A general equilibrium distribution reads
\beq
f_{\mathrm{eq}}(p) = \frac{1}{e^{\beta\left(\sqrt{p^2 + m^2}-\mu\right)} - 1} + n_0 \,\delta^{(3)}(\mathbf{p}).
\eeq
where \(\beta = 1/T\) with \(T\) the temperature, \(\mu\) is the chemical potential and \(n_0\) is the occupation density of a possible condensate. One can calculate which values of \(T\), \(\mu\) and \(n_0\) would be obtained if the system were to suddenly thermalize from the initial distribution $f_0$ to a distribution of the form $f_{\mathrm{eq}}$, with the particle density
\beq
n = \int \frac{d^3 p}{(2\pi)^3} f_0(p_{\perp},p_z) = \int \frac{d^3 p}{(2\pi)^3} f_{\mathrm{eq}}(p)
\eeq
and energy density
\beq
\epsilon = \int \frac{d^3 p}{(2\pi)^3} \sqrt{p^2 + m^2} f_0(p_{\perp},p_z) = \int \frac{d^3 p}{(2\pi)^3} \sqrt{p^2 + m^2} f_{\mathrm{eq}}(p)
\eeq
kept fixed. Note that the extracted parameters $T,\mu$ and $n_0$ depend on the time at which this sudden thermalization would happen, since a given distribution $f_0(p_\perp,\nu)$ corresponds to different distributions in $p_\perp,p_z$ at different times. Such a calculation is shown in Fig. \ref{fig:Tmu}. It assumes the occupation density in Eq. \eqref{Eq:f0} and matches it to an equilibrium distribution at various times. We see that, at all times, either \(0 < \mu \leq m\) or \(n_0 > 0\) meaning that the system is always overoccupied. A simple analysis of the occupation density factors in \(1\rightarrow 3\) and \(3\rightarrow 1\) processes shows that for an overoccupied distribution the \(3\rightarrow 1\) processes win out. Thus the net effect is to reduce the number of particles as seen in Fig. \ref{fig:number}.

\section{Summary and conclusions}
\label{sec:concl}

In this work, we have simulated quantum scalar fields in a longitudinally expanding background. This is  motivated by heavy-ion collisions where a droplet of a longitudinally expanding QCD medium is formed. A central question is how the medium becomes nearly isotropic through the competition between expansion and scatterings. To address this question, we have used the two-particle irreducible (2PI) effective action which is close to the full quantum field theory and which encompasses existing microscopic descriptions of heavy-ion collisions, namely kinetic theory and classical field theory. For simplicity, we have considered a system made of scalar fields. 

The main result is that we see signs of isotropization for the first time in a simulation of quantum fields at moderate coupling in a longitudinally expanding medium. To assess the degree of isotropization, we look at the momentum distribution of the quasiparticles and compared its $p_\perp$ and $p_z$ dependences.
While the support of the $p_z$ distribution shrinks a bit more than that of the $p_\perp$ distribution, the two remain comparable at all times, signalling the onset of isotropization. This is in contrast with what happens in the free streaming case (i.e., when we neglect the memory integrals), where the $p_z$ distribution shrinks as $\tau^{-1}$ while the $p_\perp$ distribution remains constant. 
Furthermore, we have looked at the time and momentum dependence of the thermal mass of quasiparticles. Initially, the thermal mass is highly anisotropic but then its dependences on transverse and longitudinal momenta become roughly the same, giving another sign of isotropization in our computation. Finally, we have shown the presence of number-changing processes, which may happen in the 2PI framework at order $g^4$ via off-shell processes, unlike in kinetic theory. We show that these processes lead to a roughly \(20\,\%\) drop in number density (for our overoccupied initial condition, we indeed expect a decrease of the number density).

The results presented here can be improved in several ways. Most importantly, renormalization should be done more thoroughly by including all counterterms, as well as non-Gaussian correlations in the initial state so that  the initial condition describes an interacting state instead of a non-interacting one. This would give results which have less dependency on the lattice size, allowing to quantify isotropization more accurately. Furthermore, a detailed comparison between our results on one hand and kinetic theory and classical-statistical field theory on the other hand will determine how these different approaches overlap. Ideally, such studies would use a range of different initial conditions to establish the presence of attractor solutions. 

Looking further ahead, the 2PI effective action can be used to address many questions in heavy-ion collisions. Improved renormalization would allow to calculate the stress-energy tensor and thus the ratio \(P_T / P_L\) of transverse and longitudinal pressures. This would quantify isotropization macroscopically, which would complement the microscopic approach used in this paper. Another line of study would be to consider initial conditions with a non-zero background field \(\phi\). The 2PI framework then would permit the study of the conversion of this background field into particles through parametric resonance in the presence of longitudinal expansion. This is reminiscent of the transition from a color field to quasiparticles in heavy-ion collisions. Finally, our calculation can be extended to a scalar field theory with multiple field components, truncated using a large \(N\) expansion, but with all orders in $g^2$ kept. By varying the coupling strength from weak to strong, one could then study how isotropization depends on the strength of the interactions.

\paragraph{Acknowledgements}
This work was granted access to the HPC resources of IDRIS under the allocation 2023-AD010514330 made by GENCI.
At various stages in the completion of this work, we benefited from discussions with J. Berges, J-P. Blaizot and T. Preis.

\appendix

\section{Quasi Discrete Hankel Transform}
\label{app:qdht}
In this appendix, we describe an algorithm to compute numerically the Hankel transforms that appear in the Fourier transforms of rotationally invariant functions of a 2-dimensional vector. Our method uses an approach originally developed in optics \cite{qdht1,qdht2}.

\subsection{Hankel transform with compact support}
We seek a discrete version of the Hankel transform defined in
eq.~(\ref{eq:hankel}) and its inverse, that also provides a discrete
analogue of the convolution theorem, and converges to the continuum
transform when the number of sampling points goes to infinity. Let us
assume that the function $f(r)$ has a compact support,
\begin{equation}
f(r)=0\qquad\mbox{if\ \ } r>R,
\end{equation}
and denote by $\alpha_i$ $(i=1,2,3,\cdots)$ the successive zeros of
the Bessel function $J_0(z)$. The functions $J_0(\alpha_i t)$ are
mutually orthogonal over the range $t\in[0,1]$,
\begin{equation}
  \int_0^1 dt \; t
  \;J_0(\alpha_i t)\;J_0(\alpha_k t)=
  \frac{\delta_{ik}}{2}\;J^2_1(\alpha_i) .
\end{equation}
Using this property, we see that any function $f(r)$ defined on the
range $r\in[0,R]$ can be expanded as
\begin{equation}
f(r)=\sum_{k=1}^\infty c_k\; J_0(\alpha_k\tfrac{r}{R}) ,
\label{eq:fB}
\end{equation}
with coefficients $c_k$ given by
\begin{equation}
c_k\equiv \frac{2}{R^2 J_1^2(\alpha_k)}\int_0^R dr\;r\;f(r)\;J_0(\alpha_k\tfrac{r}{R}) .
\end{equation}
Note that, up to a constant prefactor, $c_k$ is in fact the Hankel transform
of the function $f(r)$ at the momentum
$p_k\equiv\alpha_k/R$. Therefore, we define:
\begin{equation}
\tilde{f}_k\equiv\widetilde{f}(p_k) = {\pi R^2 J_1^2(\alpha_k)}\,c_k .
\end{equation}

Likewise, let us assume that the Hankel transform $\widetilde{f}(p)$
has a support $p\in[0,P]$. It can therefore be expanded as
\begin{equation}
  \widetilde{f}(p)
  =
  \sum_{i=1}^\infty d_i\;J_0(\alpha_i \tfrac{p}{P}) ,
  \label{eq:HTf}
\end{equation}
with
\begin{equation}
  d_i\equiv\frac{2}{P^2J_1^2(\alpha_i)}
  \int_0^P dp\;p\;\widetilde{f}(p)\;J_0(\alpha_i\tfrac{p}{P}) .
\end{equation}
Again we recognize, up to a prefactor, the inverse Hankel transform at
the radius $r_i\equiv \alpha_i/P$, and we define
\begin{equation}
f_i\equiv f(r_i)=\frac{P^2 J_1^2(\alpha_i)}{4\pi}\,d_i .
\end{equation}
Evaluating eq.~(\ref{eq:fB}) at $r=r_i$, and eq.~(\ref{eq:HTf}) at
$p=p_k$, we obtain the following pair of equations:
\begin{eqnarray}
  f_i=\frac{1}{\pi R^2}\sum_{k= 1}^\infty \tilde{f}_k\;
  \frac{J_0(\tfrac{\alpha_k\alpha_i}{PR})}{J_1^2(\alpha_k)}\quad,\qquad
  \tilde{f}_k=\frac{4\pi}{P^2}\sum_{i= 1}^\infty f_i\;
  \frac{J_0(\tfrac{\alpha_i\alpha_k}{PR})}{J_1^2(\alpha_i)} .
  \label{eq:DHTinf}
\end{eqnarray}
These two equations form the basis for the definition of the
discrete Hankel transform. Note that they are mutually consistent, in
the sense that the second equation exactly inverts the transformation
of the first equation.

\subsection{Finite sampling}
Equations (\ref{eq:DHTinf}) are not yet practical for numerical
implementations, since they use an infinite number of sampling points.
In order to turn these equations into a practical scheme, it is
necessary to truncate the sums on $i$ and $k$. Let us assume that we
restrict the sums to $i,k\le N_\perp $. The above equations thus
become
\begin{eqnarray}
  {f_i}
  &=&
  \frac{P}{2\pi R}\sum_{k= 1}^{N_\perp} \tilde{f}_k\;T_{ik}\;\frac{ |J_1(\alpha_i)|}{|J_1(\alpha_k)|},
  \nonumber\\
  {\tilde{f}_k}
  &=&\frac{2\pi R}{P}\sum_{i= 1}^{N_\perp}  f_i \;T_{ki}\;\frac{ |J_1(\alpha_k)|}{|J_1(\alpha_i)|}.
  \label{eq:QDHT}
\end{eqnarray}
In rewriting these formulas with a finite $N_\perp$, we have isolated
the factor $T_{ik}$, a symmetric ${N_\perp}\times {N_\perp}$ matrix
defined by
\begin{equation}
  T_{ik}\equiv
  \frac{2}{S}\frac{J_0(\tfrac{\alpha_i\alpha_k}{S})}{|J_1(\alpha_i)J_1(\alpha_k)|}\qquad(\mbox{with\ }S\equiv PR) .
\end{equation}
This makes it easy to check that Eqs.~(\ref{eq:QDHT}) are mutually
consistent provided that $T_{ik}T_{kj}=\delta_{ij}$ (with an implicit sum on the
index $k$). This property is true if $N_\perp\to +\infty$, but
violated at finite $N_\perp$.

\begin{figure}[htbp]
  \begin{center}
    \resizebox*{9cm}{!}{\includegraphics{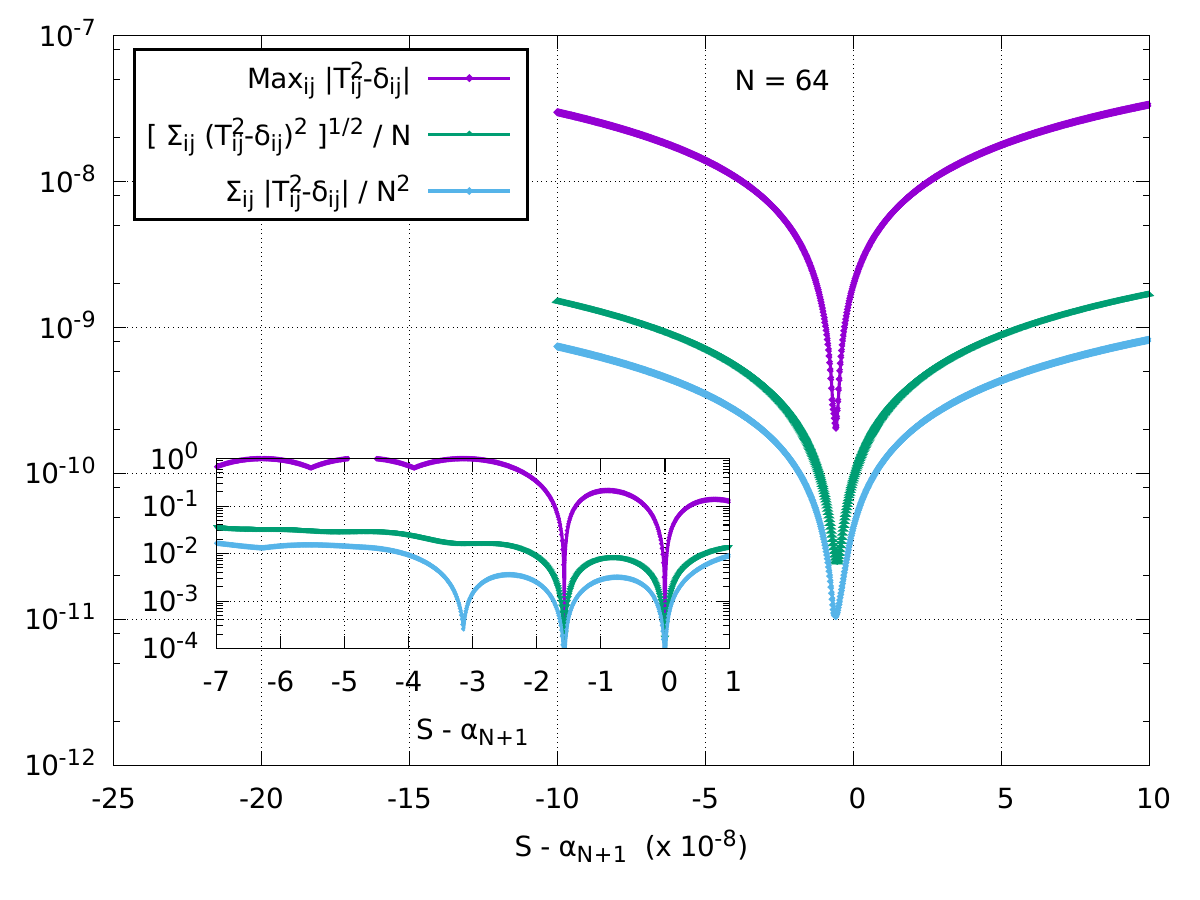}}
  \end{center}
  \caption{\label{fig:Sdep}Error on $T^2=1$ as a function of the
    parameter $S$, for $N_\perp=64$. Purple: maximal error. Green: rms
    error. Blue: average error. The insert shows the evolution of
    these errors in a larger range of values of
    $S-\alpha_{{N_\perp}+1}$, while the main plot shows these errors
    in the very close vicinity of $S\approx\alpha_{{N_{\perp}}+1}$.}
\end{figure}

However, it has been noticed (see \cite{qdht1,qdht2}) that, even for moderate values of ${N_\perp}$,
there is an optimal choice of the parameter $S\equiv PR$ such that the
matrix $T^2$ is very close to the identity. This optimal $S$ lies very
close to $\alpha_{{N_\perp}+1}$, the first zero of $J_0(z)$ not taken into
account in the truncated formulas (\ref{eq:QDHT}). We illustrate this
property in the figure \ref{fig:Sdep}, where we have studied
numerically the deviation of $T^2$ from the identity with various quantifiers for ${N_\perp}=64$ and
varying $S$.
This plot shows that there is an optimal $S_*$ such that the difference
between the coefficients of $(T^2)_{ij}$ and $\delta_{ij}$ never exceeds
$2.10^{-10}$. At this $S_*$, the rms value of $(T^2-1)_{ij}$ is of
the order of $3.10^{-11}$ and the average value of $|T^2-1|_{ij}$ is
$10^{-11}$.  Furthermore, these values decrease quickly as one
increases ${N_\perp}$, as indicated in the following table:
\begin{center}
  \begin{tabular}{c|c|c|c|c}
    ${N_\perp}$&$\alpha_{{N_\perp}+1}-S_*$& Max & RMS & Avg\\
    \hline&&&&\\
    $16$ & $3.27898.10^{-7}$ & $9.86.10^{-9}$  & $2.75.10^{-9}$  &  $1.53.10^{-9}$\\
    $32$ & $4.39245.10^{-8}$ & $1.47.10^{-9}$  & $2.84.10^{-10}$ & $1.38.10^{-10}$\\
    $64$ & $5.68755.10^{-9}$ & $1.97.10^{-10}$ & $2.46.10^{-11}$ & $1.06.10^{-11}$\\
    $128$& $7.23901.10^{-10}$& $2.53.10^{-11}$ & $1.93.10^{-12}$ & $7.49.10^{-13}$\\
    $256$& $9.14042.10^{-11}$& $3.21.10^{-12}$ & $1.43.10^{-13}$ & $5.02.10^{-14}$\\
    $512$& $1.13687.10^{-11}$& $4.21.10^{-13}$ & $1.06.10^{-14}$ & $4.33.10^{-15}$\\
  \end{tabular}
\end{center}
Beyond ${N_\perp}=512$, one may simply choose $S=\alpha_{{N_\perp}+1}$
(for instance, at ${N_\perp}=1024$, the optimal $S_*$ is
$S_*=\alpha_{{N_\perp}+1}-4.5.10^{-13}$). The pair of transformations
(\ref{eq:QDHT}), with a finite $N_\perp$, is called a Quasi Discrete
Hankel Transform (QDHT). Its implementation is straightforward, as
a matrix-vector multiplication\footnote{The cost of transforming a function represented by $N_\perp$ discrete values scales as $\sim N_\perp^2$. Therefore, unlike the usual Fast Fourier Transform that has a computational cost $\sim N_\perp \log(N_\perp)$, this algorithm is not ``fast''. Its performance is nevertheless sufficient for our needs, since we do not have to transform very large arrays.}. The coefficients that appear in the
transform, although costly to compute because they involve
Bessel functions, need only to be tabulated once.

\subsection{Basic properties}
From the definitions in eqs.~(\ref{eq:QDHT}), we can check the
following properties of the QDHT:
\begin{itemize}
\item Although the discretization does not use the point $p_\perp=0$,
  we may replace $\alpha_k/R$ by $p_k$ and formally extrapolate
  $p_k\to 0$ in the first of eqs.~(\ref{eq:qdht}) in order to obtain a
  discrete representation of $\tilde{f}(p_\perp=0)$:
  \begin{align}
    \tilde{f}(p_\perp=0)
    \empile{=}\over{\mbox{{\scriptsize discretization}}}\frac{4\pi}{P^2}\sum_{i=1}^{N_\perp} f_i\;\frac{1}{J_1^2\big(\alpha_i\big)}.
  \end{align}
  The r.h.s of the previous equation can therefore be viewed as the
  discretization of the radial integral of the function $f(r_\perp)$:
  \begin{align}
    2\pi \int_0^{+\infty}
    dr_\perp\,r_\perp\,f(r_\perp)
    \empile{=}\over{\mbox{{\scriptsize discretization}}}
    \frac{4\pi}{P^2}\sum_{i=1}^{N_\perp} f_i\;\frac{1}{J_1^2\big(\alpha_i\big)}.
  \end{align}
  \item Likewise, by extrapolating $\alpha_i/P=r_i\to 0$ in the second
    of eqs.~(\ref{eq:qdht}), we get
    \begin{align}
      f(r_\perp=0)=\int_0^{+\infty}\frac{dp_\perp\,p_\perp}{2\pi}\,\tilde{f}(p_\perp)
    \empile{=}\over{\mbox{{\scriptsize discretization}}}\frac{1}{\pi R^2}\sum_{k=1}^{N_\perp} \tilde{f}_k\;\frac{1}{J_1^2\big(\alpha_k\big)}.
    \end{align}
  \item We may also replace $\alpha_i/P$ by a continuously varying
    $r_\perp$ in the second of eqs.~(\ref{eq:qdht}), and then take the
    transverse Laplacian. Using the Bessel equation obeyed by $J_0$,
    we obtain
    \begin{align}
      {\big(\Delta_\perp f\big)}_i&=\frac{1}{\pi R^2}\sum_{i=1}^{N_\perp} \big(-p_k^2\big)\tilde{f}_k\;\frac{J_0\big(\tfrac{\alpha_i\alpha_k}{PR}\big)}{J_1^2\big(\alpha_i\big)},
    \end{align}
    which is consistent with the fact that the Fourier transform of
    $\Delta_\perp f(r_\perp)$ is $-p_\perp^2 \tilde{f}(p_\perp)$.
\end{itemize}

\subsection{Additional properties valid when $T^2\approx 1$}
When $T_{ik}T_{k j}\approx \delta_{ij}$, a number of standard properties of Fourier transforms
become true with the same accuracy between the discretized quantities:
\begin{itemize}
  \item{\bf Orthogonality of Bessel functions:} the property of orthogonality of Bessel functions,
    \begin{align}
      \int_0^{+\infty} dr_\perp\, r_\perp\, J_0(p_\perp r_\perp) J_0(q_\perp r_\perp)
      =
      \frac{1}{p_\perp}\delta(p_\perp-q_\perp),
    \end{align}
    has the following discrete counterpart:
    \begin{align}
      \frac{2}{P^2}\sum_{i=1}^{N_\perp}\frac{J_0\big(\tfrac{\alpha_i\alpha_p}{PR}\big)J_0\big(\tfrac{\alpha_i\alpha_q}{PR}\big)}{J_1^2(\alpha_i)}
      \approx
      \frac{R^2}{2} \delta_{pq} J_1^2(\alpha_p).
    \end{align}
  (The equation is in fact equivalent to the statement $T^2\approx
    1$.)
\item{\bf Convolution theorem:} given two functions $f(r_\perp)$ and
  $g(r_\perp)$, we define
  \begin{align}
    h(r_\perp)\equiv \int d^2\u_\perp\,f(u_\perp)g(|\r_\perp-\u_\perp|).
  \end{align}
  Their respective discretizations satisfy a property which is an
  analogue of the convolution theorem:
  \begin{align}
    h_i \approx \frac{1}{\pi R^2}\sum_{i=1}^{N_\perp} \tilde{f}_k \tilde{g}_k\; \frac{J_0\big(\tfrac{\alpha_i\alpha_k}{PR}\big)}{J_1^2\big(\alpha_i\big)}.
  \end{align}
  \item{\bf Parseval theorem:} given two functions $f(r_\perp)$ and
    $g(r_\perp)$, one may check that:
    \begin{align}
      \frac{4\pi}{P^2}\sum_{i=1}^{N_\perp}
      \frac{f_i g_i}{J_1^2(\alpha_i)}
      \approx
      \frac{1}{\pi R^2}
      \sum_{k=1}^{N_\perp} \frac{\tilde{f}_k\tilde{g}_k}{J_1^2(\alpha_k)},
    \end{align}
    again with an error of the order of $T^2-1$.
\end{itemize}

\section{Properties of Hankel functions}
We collect in this appendix some useful formulas about the Hankel
functions $H_{i\nu}^{(1,2)}(z)$. These functions form a basis of
the two-dimensional space of solutions of the following Bessel
equation:
\begin{align}
  \partial_z^2 X+\frac{1}{z}\partial_z X + X + \frac{\nu^2}{z^2} X=0.
\end{align}
Note that an alternate basis is provided by Bessel functions of the first and second kind, $J_{i\nu}(z)$ and
$Y_{i\nu}(z)$. The two sets of functions are related by
\begin{align}
  H_{i\nu}^{(1)}(z)\equiv J_{i\nu}(z)+i Y_{i\nu}(z),\quad
    H_{i\nu}^{(2)}(z)\equiv J_{i\nu}(z)-i Y_{i\nu}(z).
\end{align}

The Hankel functions have the following Taylor expansion at small $z$:
\begin{align}
  &H_{i\nu}^{(1)}(z)\empile{=}\over{z\approx 0}
  \frac{1}{\Gamma(1+i\nu)}\Big(\frac{z}{2}\Big)^{i\nu}
  \Big(1+\frac{\cosh(\pi\nu)}{\sinh(\pi\nu)}\Big)
  -\frac{1}{\Gamma(1-i\nu)}\Big(\frac{z}{2}\Big)^{-i\nu}+{\cal O}(z^{2\pm i\nu}),\nonumber\\
  &H_{i\nu}^{(2)}(z)\empile{=}\over{z\approx 0}
   \frac{1}{\Gamma(1+i\nu)}\Big(\frac{z}{2}\Big)^{i\nu}
  \Big(1-\frac{\cosh(\pi\nu)}{\sinh(\pi\nu)}\Big)
  +\frac{1}{\Gamma(1-i\nu)}\Big(\frac{z}{2}\Big)^{-i\nu}+{\cal O}(z^{2\pm i \nu})
  ,
\end{align}
and the following asymptotic behavior at large $z$:
\begin{align}
  H_{i\nu}^{(1)}(z)\empile{\approx}\over{z\to +\infty}
  \sqrt{\frac{2}{\pi z}}\;
  e^{+i(z-i\tfrac{\pi \nu}{2}-\tfrac{\pi}{4})},\quad
  H_{i\nu}^{(2)}(z)\empile{\approx}\over{z\to +\infty}
  \sqrt{\frac{2}{\pi z}}\;
  e^{-i(z-i\tfrac{\pi \nu}{2}-\tfrac{\pi}{4})}.
\end{align}
The latter relations show that the Hankel functions generalize the
concept of positive and negative frequency solutions to the case of
the Bessel equation.

They admit several integral representations. In the context of the
present paper, a particularly useful one is
\begin{align}
  H_{i\nu}^{(1)}(z)
  &=
  \frac{e^{\pi\nu/2}}{i\pi}\int_{-\infty}^{+\infty} dt\;e^{i(z\cosh(t)-\nu t)},\nonumber\\
  \quad
  H_{i\nu}^{(2)}(z)
  &=
  -\frac{e^{-\pi\nu/2}}{i\pi}\int_{-\infty}^{+\infty} dt\;e^{-i(z\cosh(t)+\nu t)}.
  \label{eq:Hint}
\end{align}
From these integral representations, we see that
\begin{align}
  H_{i\nu}^{(2)}(z)=e^{-\pi \nu} \Big(H_{i\nu}^{(1)}(z)\Big)^*\quad\mbox{for $z$ and $\nu$ real.}
\end{align}
Finally, let us mention the Wronskian, which is a simple invariant of
the evolution in $z$:
\begin{align}
H_{i\nu}^{(1)}(z)\dot{H}_{i\nu}^{(2)}(z)-\dot{H}_{i\nu}^{(1)}(z)H_{i\nu}^{(2)}(z)=\frac{4}{i\pi z}.
\end{align}

\section{Bare propagators in Milne coordinates}
\label{app:bareprop}
In this appendix, we derive the pro\-pa\-ga\-tors
$F_0(\tau,\tau';p_\perp,\nu)$ and $\rho_0(\tau,\tau';p_\perp,\nu)$ of the
non-interacting theory. Thanks to eqs.~(\ref{eqs:Frho}), this is
equivalent to obtaining the propagators $G_{-+}$ and $G_{+-}$.

Let us first derive the bare propagators in vacuum. Recall that,
with Cartesian coordinates and momenta, these propagators are given
 by
\begin{align}
  G_{+-}^{0\;\rm vac}(x_0,x_0';\p)&=\frac{1}{2E_\p}e^{+iE_\p(x_0-x_0')},\nonumber\\
  G_{-+}^{0\;\rm vac}(x_0,x_0';\p)&=\frac{1}{2E_\p}e^{-iE_\p(x_0-x_0')}.
\end{align}
Note that $G_{+-}^{0\;\rm vac}$ has positive frequency oscillations in
$x_0$ (and a negative frequency in $x_0'$), while $G_{-+}^{0\; \rm vac}$
has negative frequency oscillations in $x_0$ (and a positive frequency
in $x_0'$). In Milne coordinates, $G_{+-}^{0\;\rm
  vac}(\tau,\tau';p_\perp,\nu)$ and $G_{-+}^{0\;\rm
  vac}(\tau,\tau';p_\perp,\nu)$ both satisfy the Bessel equation
\begin{align}
\Big(\partial_\tau^2+\frac{1}{\tau}\partial_\tau+m^2+p_\perp^2+\frac{\nu^2}{\tau^2}\Big)X(\tau)=0,
\end{align}
and the same equation for their $\tau'$ dependence. The two Hankel
functions $H_{i\nu}^{(1,2)}(m_\perp\tau)$ (with $m_\perp\equiv
\sqrt{p_\perp^2+m^2}$) provide a basis of the space of solutions of
this equation. Therefore, we may write $X(\tau$ as
\begin{align}
  X(\tau)= \alpha_1 H_{i\nu}^{(1)}(m_\perp\tau)+\alpha_2 H_{i\nu}^{(2)}(m_\perp\tau).
\end{align}
Thanks to the integral representations in eq.~(\ref{eq:Hint}),
we see that $H_{i\nu}^{(1)}(m_\perp\tau)$ has positive frequency
oscillations in $\tau$, while $H_{i\nu}^{(2)}(m_\perp\tau)$ has
negative frequency oscillations. This implies that
\begin{align}
  G_{+-}^{0\;\rm vac}(\tau,\tau';p_\perp,\nu)&\propto H_{i\nu}^{(1)}(m_\perp\tau)H_{i\nu}^{(2)}(m_\perp\tau'),\nonumber\\
  G_{-+}^{0\;\rm vac}(\tau,\tau';p_\perp,\nu)&\propto H_{i\nu}^{(2)}(m_\perp\tau)H_{i\nu}^{(1)}(m_\perp\tau').
\end{align}
The missing prefactor can be found by considering the spectral
function\footnote{We can omit the superscript ``vac'' for the spectral
function, because its bare value does not depend on the presence of an
ensemble of particles.} $\rho_0\equiv -i (G_{-+}^{0\; \rm vac}-G_{+-}^{0\;\rm
  vac})$,
\begin{align}
  \rho_0(\tau,\tau';p_\perp,\nu)\propto \Big(H_{i\nu}^{(2)}(m_\perp\tau)H_{i\nu}^{(1)}(m_\perp\tau')- H_{i\nu}^{(1)}(m_\perp\tau)H_{i\nu}^{(2)}(m_\perp\tau')\Big).
\end{align}
On the other hand, from $\rho(x,x')=-i\big<[\phi(x),\phi(x')]\big>$ and the
canonical commutation relation of the field and its conjugate
momentum, we have
\begin{align}
\partial_{x_0'}\rho_0(x,x')\empile{=}\over{x_0=x_0'}\delta(\x-\x'),
\end{align}
from which we get
\begin{align}
\partial_{\tau'}\rho_0(\tau,\tau'; p_\perp,\nu)\empile{=}\over{\tau=\tau'}\frac{1}{\tau}\qquad(\mbox{and\ \ } \partial_{\tau}\rho_0(\tau,\tau'; p_\perp,\nu)\empile{=}\over{\tau=\tau'}-\frac{1}{\tau}),
\end{align}
Using the Wronskian of the Hankel functions, we find that \begin{align}
  \rho_0(\tau,\tau';p_\perp,\nu)=\frac{i\pi}{4}\Big(H_{i\nu}^{(1)}(m_\perp\tau)H_{i\nu}^{(2)}(m_\perp\tau')- H_{i\nu}^{(2)}(m_\perp\tau)H_{i\nu}^{(1)}(m_\perp\tau')\Big).
\end{align}
From this, we obtain the prefactor in $G_{+-}^{0\;\rm vac}$ and $G_{-+}^{0\;\rm vac}$,
\begin{align}
  G_{+-}^{0\;\rm vac}(\tau,\tau';p_\perp,\nu)&=\frac{\pi}{4} H_{i\nu}^{(1)}(m_\perp\tau)H_{i\nu}^{(2)}(m_\perp\tau'),\nonumber\\
  G_{-+}^{0\;\rm vac}(\tau,\tau';p_\perp,\nu)&=\frac{\pi}{4} H_{i\nu}^{(2)}(m_\perp\tau)H_{i\nu}^{(1)}(m_\perp\tau'),
\end{align}
as well as the statistical propagator
\begin{align}
  F^{0\;\rm vac}(\tau,\tau';p_\perp,\nu)=\frac{\pi}{8} \Big\{
  H_{i\nu}^{(1)}(m_\perp\tau)H_{i\nu}^{(2)}(m_\perp\tau')
  +
  H_{i\nu}^{(2)}(m_\perp\tau)H_{i\nu}^{(1)}(m_\perp\tau')\Big\}.
\end{align}
When we include a particle  distribution $f(p_\perp,\nu; \tau)$, $\rho$ is
not modified and, in the case of $F$, a prefactor $\tfrac{1}{2}$ should
be changed into $\tfrac{1}{2}+f(p_\perp,\nu; \tau)$,
\begin{align}
  F_0(\tau,\tau';p_\perp,\nu)=\frac{\pi\big(\tfrac{1}{2}+f(p_\perp,\nu; \tau)\big)}{4}& \Big\{
  H_{i\nu}^{(1)}(m_\perp\tau)H_{i\nu}^{(2)}(m_\perp\tau')
  \nonumber\\
  &+
  H_{i\nu}^{(2)}(m_\perp\tau)H_{i\nu}^{(1)}(m_\perp\tau')\Big\}.
\end{align}

\section{Quasi-particle approximation}
\label{App:quasiparticle}
\subsection{Propagators in the quasi-particle approximation}
A system that admits a quasi-particle interpretation is one in which
the propagators $F$ and $\rho$ take the same forms as the free ones,
\begin{align}
  {F}(\tau,\tau';p_\perp,\nu)
  &=
  \frac{\pi (\tfrac{1}{2}+f(p_\perp,\nu;\tau))}{4}
  \!\Big(\!
  H_{i\nu}^{(1)}(m_\perp \tau)H_{i\nu}^{(2)}(m_\perp \tau')
  \nonumber\\
  &\qquad\qquad\qquad\qquad+\!
  H_{i\nu}^{(2)}(m_\perp \tau)H_{i\nu}^{(1)}(m_\perp \tau')
  \!\Big),
\end{align}
\begin{align}
  {\rho}(\tau,\tau';p_\perp,\nu)
  &=
  \frac{i\pi}{4}
  \!\Big(\!
  H_{i\nu}^{(1)}(m_\perp \tau)H_{i\nu}^{(2)}(m_\perp \tau')
  \!-\!
  H_{i\nu}^{(2)}(m_\perp \tau)H_{i\nu}^{(1)}(m_\perp \tau')
  \!\Big).
\end{align}
In these expressions of the propagators, the transverse mass
$m_\perp(\tau)\equiv \sqrt{p_\perp^2+m^2+M^2(\tau)}$ may contain a mass correction $M(\tau)$ due to in-medium effects and vacuum fluctuations. 

\subsection{Extraction of the occupation number}
The implicit assumption in a quasi-particle description is that the
occupation number $f(p_\perp,\nu; \tau)$, although it may be time dependent,
has a slow time dependence compared to the timescale of the
oscillations of the Hankel functions. The same assumption is made
about the mass, that can become medium-dependent (and thus
time-dependent) due to the interactions of particles with the
surrounding medium.  Therefore, in the time derivatives of the
propagators, we may neglect the terms containing the derivative of the
occupation number or of the mass as in, e.g.,
\begin{align}
   \partial_\tau{F}(\tau,\tau';p_\perp,\nu)
  \approx
  \frac{\pi m_\perp  (\tfrac{1}{2}+f(p_\perp,\nu; \tau))}{4}
  \!&\Big(\!
  \dot{H}_{i\nu}^{(1)}(m_\perp \tau)H_{i\nu}^{(2)}(m_\perp \tau')
  \nonumber\\
  &+\!
  \dot{H}_{i\nu}^{(2)}(m_\perp \tau)H_{i\nu}^{(1)}(m_\perp \tau')
  \!\Big).
\end{align}
Using the Wronskian, this approximation leads to the following relationship
\begin{align}
  \tau\sqrt{F(\partial_\tau\partial_{\tau'}F)-(\partial_\tau F)(\partial_{\tau'}F)}\empile{\approx}\over{\tau=\tau'} \frac{1}{2}+f(p_\perp,\nu; \tau).
\end{align}
The advantage of this formula is that it allows to extract the
occupation number without prior knowledge of the mass $M$ that enters
in $m_\perp$, which is especially useful when in-medium effects change
dynamically the effective mass of the quasi-particles.

\subsection{Extraction of the transverse mass}
Given the propagators $F(\tau,\tau';p_\perp,\nu)$ and
$\rho(\tau,\tau';p_\perp,\nu)$ (and the time derivatives of $F$ at
$\tau=\tau'$, in order to first obtain the occupation number), it is
also possible to extract the transverse mass $m_\perp$, assuming the
quasi-particle approximation. Firstly, one should determine the
occupation number by the method outlined above.  Then, one may
consider the following combination:
\begin{align}
  {\bm{\Upgamma}}(\tau,\tau';p_\perp,\nu)
  \equiv
  \frac{F(\tau,\tau';p_\perp,\nu)}{\tfrac{1}{2}+f(p_\perp,\nu)}-i\rho(\tau,\tau';p_\perp,\nu)
  \approx
  \frac{\pi}{2}\,H_{i\nu}^{(1)}(m_\perp \tau) H_{i\nu}^{(2)}(m_\perp \tau'), 
\end{align}
where the expression in the right hand side results from the
quasi-particle approximation. By decreasing $\tau'$ at fixed $\tau$,
starting from $\tau'=\tau$, we determine the oscillation period of the
phase\footnote{In the system of expanding Milne coordinates, this
quantity is not periodic in $\tau'$. But its value revolves around
$z=0$ in the complex plane, and one may define a pseudo-period as
the minimal variation of $\tau'$ necessary for one revolution around $z=0$.}
of this quantity:
\begin{align}
  T(\tau;p_\perp,\nu)\equiv
   \min_{\tau'<\tau}\Big\{\tau-\tau'\Big|{\rm Phase}\,
     ({\bm{\Upgamma}}(\tau,\tau';p_\perp,\nu))={\rm Phase}\,
    ({\bm{\Upgamma}}(\tau,\tau;p_\perp,\nu))\Big\}.
\end{align}
We define similarly
the period of the phase oscillations of  the Hankel function $H_{i\nu}^{(2)}(x)$:
\begin{align}
  T_{_H}(x;\nu)\equiv
  \min_{x'<x}\Big\{x-x'\Big|{\rm Phase}\,(H_{i\nu}^{(2)}(x'))={\rm Phase}\,(H_{i\nu}^{(2)}(x))\Big\}.
\end{align}
The transverse mass $m_\perp$ satisfies the following equation,
\begin{align}
  m_\perp T(\tau;p_\perp,\nu)=T_{_H}(m_\perp\tau;\nu).
\end{align}
This is a non-linear equation for $m_\perp$, that can be solved
numerically by the following iterative algorithm:
\begin{align}
  &  m_\perp^{(0)}=\mbox{some initial value smaller that $m$},\nonumber\\
  &  m_\perp^{(i+1)} = \frac{T_{_H}(m_\perp^{(i)}\tau;\nu)}{T(\tau;p_\perp,\nu)}.
\end{align}
From the transverse mass, one may obtain the correction $M$ to the mass by
\begin{align}
M^2\equiv m_\perp^2-p_\perp^2-m^2.
\end{align}
Note that this effective mass may depend on $p_\perp$ and $\nu$ when a
momentum dependent self-energy is included in the right hand side of
the equations of motion for the propagators.

\bibliography{refs}
\bibliographystyle{jhep}

\end{document}